\RequirePackage{amsthm}

\documentclass[sn-mathphys,Numbered, iicol]{sn-jnl}

\usepackage{graphicx}%
\usepackage{multirow}%
\usepackage{amsmath,amssymb,amsfonts}%
\usepackage{mathrsfs}%
\usepackage[title]{appendix}%
\usepackage{xcolor}%
\usepackage{textcomp}%
\usepackage{manyfoot}%
\usepackage{booktabs}%
\usepackage{algorithm}%
\usepackage{algorithmicx}%
\usepackage{algpseudocode}%
\usepackage{listings}%
\usepackage{physics}
\usepackage{calc}
\usepackage{cleveref}

\lstdefinestyle{customc}{
  belowcaptionskip=1\baselineskip,
  breaklines=true,
  frame=single,
  xleftmargin=\parindent,
  language=C,
  showstringspaces=false,
  basicstyle=\ttfamily,
  keywordstyle=\bfseries\color{green!40!black},
  commentstyle=\itshape\color{purple!40!black},
  identifierstyle=\color{black}, 
  stringstyle=\color{orange},
}

\theoremstyle{thmstyleone}%
\theoremstyle{thmstyletwo}%

\theoremstyle{thmstylethree}%

\newlength{\firsttermlength}
\providecommand{\firstterm}{}
\newcommand{\ftspace}{\hspace{\firsttermlength}}

\newcommand{\setfirstterm}[1]{%
\renewcommand{\firstterm}{\ensuremath{#1}}%
\setlength{\firsttermlength}{\widthof{\firstterm}}}

\begin{document}

\title[Article Title]{\centering \textbf{KinFit} \\ A Kinematic Fitting Package for Hadron Physics Experiments}


\author[1]{\fnm{Waleed} \sur{Esmail}}\email{w.esmail@gsi.de}

\author[2]{\fnm{Jana} \sur{Rieger}}\email{jana.rieger@physics.uu.se}

\author*[1,2]{\fnm{Jenny} \sur{Taylor}}\email{j.taylor@gsi.de}

\author[2]{\fnm{Malin} \sur{Bohman}}

\author[2]{\fnm{Karin} \sur{Schönning}}

\affil[1]{\orgdiv{GSI-FFN}, \orgname{GSI Helmholtzzentrum fur Schwerionenforschung GmbH}, \orgaddress{\street{Planckstraße~1}, \city{Darmstadt}, \postcode{642 91}, \state{Hesse}, \country{Germany}}}

\affil[2]{\orgdiv{Department of Physics and Astronomy}, \orgname{Uppsala~University}, \orgaddress{\street{Lägerhyddsvagen~1}, \city{Uppsala}, \postcode{752 37}, \country{Sweden}}}

\abstract{A kinematic fitting package, \lstinline|KinFit|, based on the Lagrange multiplier technique has been implemented for generic hadron physics experiments. It is particularly suitable for experiments where the interaction point is unknown, such as experiments with extended target volumes. The \lstinline|KinFit| package includes vertex finding tools and fitting with kinematic constraints, such as mass hypothesis and four-momentum conservation, as well as combinations of these constraints. The new package is distributed as an open source software via GitHub. 

This paper presents a comprehensive description of the KinFit package and its features, as well as a benchmark study using Monte Carlo simulations of the $pp\rightarrow pK^+\Lambda \rightarrow pK^+p\pi^-$ reaction. The results show that  \lstinline|KinFit| improves the parameter resolution and provides an excellent basis for event selection.}


\maketitle
\section{Introduction}
\label{sec:introduction}

Kinematic fitting is a powerful tool widely used in particle and nuclear physics analyses, recognised for its ability to improve the resolution of measured particle track parameters, suppress background, reconstruct undetected particles and to determine the position of vertices. Available information from measurements, such as momenta, angles and energy, combined with physics constraints, such as four-momentum conservation in a production or displaced decay vertex, or the mass of an undetected unstable particle through its decay products, are exploited. With this information, a mathematical minimization problem can be formulated and solved using \textit{e.g.}, the Lagrange multiplier technique.

 Kinematic fitting techniques have been available in particle physics since the 1960's~\cite{Bock:278580} and gained momentum during the bubble chamber days. However, existing packages like RAVE~\cite{RAVE} from ILD or the full decay chain fitters from Belle II~\cite{belle} or PANDA~\cite{panda} are often embedded in the respective experiment software and specialized for the detector setup. In the measurement of complex decays of heavy particles, it is beneficial apply kinematic tree fits ~\cite{belle} to complex decay chains or jets~\cite{RAVE, https://doi.org/10.48550/arxiv.2110.13731}. 

\begin{figure}
    \centering
    \includegraphics[width=0.5\textwidth]{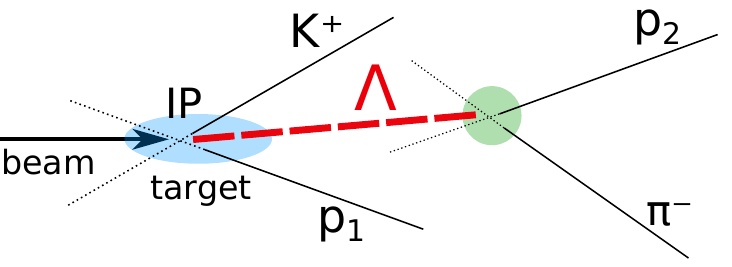}
    \caption{Illustration of a $\Lambda$ hyperon produced by a proton beam hitting a proton target. The position of the interaction point (IP, marked in blue) and the $\Lambda$ decay vertex (marked in green) are determined by the point of closest approach of the $p_1$ and $K$ and $p_2$ and $\pi^-$ tracks, respectively.}
    \label{fig:lambda_vertices}
\end{figure}

In many hadron physics experiments like the HADES experiment~\cite{Agakichiev_2009} at GSI, the interaction point is not fixed to a point but can be located within a target volume that covers several centimeters of the path along the beamline. Hence, the interaction vertex position needs to be determined from the measured track parameters of the outgoing particles. In addition, weakly decaying particles such as hyperons produce daughter particles in a secondary vertex, located a measurable distance away from the interaction point. These are crucial to reconstruct since many hyperons, \textit{e.g.} $\Lambda$ and $\Sigma^0$, are neutral and could otherwise escape detection. Figure~\ref{fig:lambda_vertices} illustrates an example where a $\Lambda$ hyperon is produced in the target volume and subsequently decays in a secondary vertex after travelling some distance. Tracks from secondary particles are more challenging to reconstruct and hence, the reconstructed track parameters have larger measurement uncertainties compared to those from primary particles. All this calls for kinematic fitting. A new package, \lstinline|KinFit|, has been developed to facilitate the hyperon program at HADES~\cite{HADES:2020pcx} but is provided in an experiment-agnostic way and is therefore applicable also to other experiments. Hadron physics experiments at J-PARC~\cite{J-PARC}, JLab~\cite{JLab} or COMPASS~\cite{compass} and AMBER at CERN, may profit from this package.

In addition to kinematic fitting, tools are provided to construct a particle candidate from its daughter tracks. A functionality for running a kinematic fit for event selection on a ROOT~\cite{root} file in an automated way is included as well. 

This paper is outlined as follows: The methodology including the fitting procedure and available constraint equations are described in Section~\ref{sec:method}. In Section~\ref{sec:classes}, the classes contained in the package are described and in Section~\ref{sec:performance}, a benchmark study is provided for $\Lambda$ hyperon reconstruction to demonstrate the performance of the fitter. In addition, a comprehensive user's guide is provided in \Cref{app:usersguide}. 

\section{Methodology} \label{sec:method}
The goal of a kinematic fitting algorithm is to find an improved set of track parameters as close as possible to the true values, such that they fulfill a set of kinematic and/or geometric constraints. This concept can be translated quantitatively into a $\chi^{2}$ minimization expressed as follows:
\begin{equation}\label{eq:chi2}
    \chi^{2} = (\vec{y} - \vec{\eta})^{T} \: V^{-1}(\vec{y})\: (\vec{y} - \vec{\eta}) = \mathrm{minimum} \:,
\end{equation}

\noindent
where  $\vec{y}$ is a vector of the $N\in\mathbb{N}$ measured track parameters provided by the tracking algorithm, $V(\vec{y})$ is the corresponding covariance matrix and $\vec{\eta}$ are the estimated track parameters.

In addition, the kinematic and geometric constraints are implemented in the procedure by constraint equations $\vec{f}$. These are in general represented by a set of $K\in\mathbb{N}$ continuously differentiable functions of the estimated parameters $\vec{\eta}$ and a set of $J\in\mathbb{N}_0$ unmeasured parameters combined in $\vec{\xi}$:

\begin{equation*}
    \vec{f} = \vec{f}(\vec{\eta},\,\vec{\xi}) = 0
\end{equation*}
\\
The objective is to minimize the $\chi^2$ from \Cref{eq:chi2} subject to these constraints. 

\subsection{The Iterative Lagrange Multiplier Method}
The method of Lagrange multipliers \cite{Frodesen:1979fy} is well-suited for addressing such problems. This technique enables the transformation of the constrained minimization into the minimization of a single \textit{Lagrange function}, $\mathcal{L}$:

\begin{align} \label{eq:2.1}
    \mathcal{L} &= (\vec{y} - \vec{\eta})^{T}V^{-1}(\vec{y} - \vec{\eta}) + 2 \vec{\lambda}^{T}f(\vec{\eta}, \vec{\xi}) \:,
\end{align}

\noindent
where the $K$ additional variables summarized in $\vec{\lambda}$ are introduced, referred to as \textit{Lagrange multipliers}.

Minimizing the Lagrange function (\Cref{eq:2.1}) involves finding the derivatives of $\mathcal{L}$ with respect to all unknowns $\vec{\eta}$, $\vec{\xi}$, $\vec{\lambda}$. By setting these derivatives to zero and subsequently solving for $\vec{\eta}$ and $\vec{\xi}$, the minimization can be achieved. As this problem is generally non-linear, an iterative procedure is employed, with each iteration yielding improved approximations for $\vec{\eta}$ and $\vec{\xi}$. Assuming the values of all quantities of the iteration $\nu$ have already been extracted, we want to express the quantities of the subsequent iteration $\nu + 1$ in terms of the values of iteration $\nu$. We then proceed as follows \cite{Frodesen:1979fy}:

\begin{enumerate}
    \item First, the following notations are introduced 
    \begin{align*}
        \vec{r} &= \vec{f}^{\nu} + F^{\nu}_{\eta} (\vec{y} - \vec{\eta}^{\nu}) \\
        S &= F^{\nu}_{\eta}V(F^{\nu}_{\eta})^{T} \:,
    \end{align*}
    
\noindent where $F_{\eta}$ is a $K \times N$ Jacobian matrix ($F_{\eta}=D_\eta\vec{f}$), \textit{i.e.} the derivative of the constraint equations with respect to $\vec{\eta}$.

    \item The updated, unmeasured variables, $\vec{\xi}^{\nu + 1}$, are obtained by
    \begin{align*}
        \vec{\xi}^{\nu + 1} = \vec{\xi}^{\nu} - (F_{\xi}^{T}S^{-1}F_{\xi})^{-1}F_{\xi}^{T}S^{-1}\vec{r} \:,
    \end{align*}
    
\noindent where $F_{\xi}$ is a $K \times J$ Jacobian matrix ($F_{\xi} = D_\xi\vec{f}$), \textit{i.e.} the derivative of the constraint equations with respect to $\vec{\xi}$.  
    
    \item The updated Lagrange multipliers, $\vec{\lambda}^{\nu + 1}$, are obtained from 
    \begin{align*}
        \vec{\lambda}^{\nu +1} = S^{-1} \Big( \vec{r} + F_{\xi}(\vec{\xi}^{\nu +1}-\vec{\xi}^{\nu}) \Big) \:.
    \end{align*}
    
    \item The updated measured parameters, $\vec{\eta}^{\nu + 1}$, are calculated as
    \begin{align*}
        \vec{\eta}^{\nu +1} = \vec{y} - VF_{\eta}^{T}\vec{\lambda}^{\nu +1} \:.
    \end{align*}
    
    \item Finally, the new $\mathcal{L}$ is calculated and the results compared with the previous iteration. To decide when the solution is sufficiently close to the minimum, convergence criteria are defined, see \Cref{subsec:convergence}.
    
\end{enumerate}

In the end, the new covariance matrix, $V^{\nu + 1}$, is calculated
\setfirstterm{V^{\nu +1}}
\begin{multline}
\firstterm\:=\:V - V  [ F^{T}_{\eta}S^{-1}F_{\eta} \\ 
\:-\:(F^{T}_{\eta}S^{-1}F_{\xi})(F^{T}_{\xi}S^{-1}F_{\xi})^{-1}(F^{T}_{\eta}S^{-1}F_{\xi})^{T} ] V \:.
\end{multline}

\noindent This ansatz assumes a quadratic minimum of the $\chi^2$ function. The function will look different if it deviates too much from its minimum. Currently there is no guard against this but the user should be mindful of potential deviations and assess the validity of the results accordingly.

\subsection{Track Parametrization}\label{sec:trackparametrization}

The input objects for \lstinline|KinFit| include track parameters such as momentum and angles, defined in polar coordinates, as well as the point of closest approach to the beam axis. Functions for converting from Cartesian coordinates are provided to ensure compatibility with other experiments (see the User's Guide in Appendix \ref{app:usersguide}). In addition, the covariance matrix is required as input. The particles are assumed to move in a region free of magnetic fields, meaning they are propagated as straight tracks. The parameters employed here to uniquely describe a track include: 

\begin{itemize}
	\item Inverse momentum $1/p$ [MeV$^{-1}$\textit{c}]. 
	\item Polar angle $\theta$ [radians] defined from $0$ to $\pi$.
	\item Azimuthal angle $\phi$ [radians] defined from $-\pi$ to $\pi$ relative to the beam ($z$) axis.
	\item $R$ [mm], the distance between the beam axis and the point of closest approach of the track to the beam axis.
	\item $Z$ [mm], the $z$ coordinate of the point of closest approach of the track to the beam axis.
    \vspace{5pt}
    \\
	and
    \vspace{5pt}
	\item Covariance matrix, where the diagonal entries correspond to the uncertainties in the parameters $1/p$, $\theta$, $\phi$, $R$ and $Z$. 
\end{itemize}

For the purely kinematic fits, only the track parameters $1/p$, $\theta$ and $\phi$ are required.

\subsection{Constraints}\label{sec:constraints}
\lstinline|KinFit| offers a variety of constraints that can be selected for kinematic fitting.

\subsubsection{1C: Vertex Constraint}
\label{sec:vertex}
This purely geometrical vertex constraint minimizes the distance between two tracks, ensuring they originate from the same vertex. A straight line in 3D is uniquely defined by a base vector that points from the origin of a chosen coordinate system to a coordinate on the line and the direction of the line. The components of these vectors are:

\begin{equation}
\begin{cases}
b_x=R\cdot\cos(\phi+\pi/2), \\
b_y=R\cdot\sin(\phi+\pi/2), \\
b_z=Z,
\end{cases}
\end{equation}
and

\begin{equation}
\begin{cases}
d_x=\sin(\theta)\cdot\cos(\phi), \\
d_y=\sin(\theta)\cdot\sin(\phi), \\
d_z=\cos(\theta).
\end{cases}
\end{equation}
\ \\
In the base vector, $\pi/2$ is added to $\phi$ to ensure the base vector is constructed in the HADES coordinate system. A suitable constraint equation can then be formulated as follows: 

\begin{equation}
    f=(\vec{d_1}\times\vec{d_2})\cdot(\vec{b_1}-\vec{b_2}) = 0.
\end{equation}

This expression is proportional to the minimum distance between the straight lines parameterized by the respective base and direction vectors $\Vec{b_1}$, $\Vec{b_2}$, $\Vec{d_1}$ and $\Vec{d_2}$. 

The vertex fit can be performed either separately or as part of a series of consecutive fitting procedures. In addition, there is the possibility to perform a mass and a geometrical vertex fit simultaneously.

\subsubsection{4C: Four-momentum conservation in the beam-target interaction}
\label{sec:4C}
The 4C constraint demands that the sum of the four-momenta of the final state particles equals that of the initial beam-target system (in the following denoted with the suffix \texttt{ini}). Since all parameters are treated as measured parameters with uncertainties, there are four over-constraints. For $N$ particles, the constraint equations are

\begin{equation}
f =
\begin{cases}

\sum\limits^N_{i=1} p_{i}\sin\theta_{i}\cos\phi_{i}-p_{\mathrm{ini}, x}=0 \text{ }\text{ } (p_x), \\

\sum\limits^N_{i=1} p_{i}\sin\theta_{i}\sin\phi_{i}-p_{\mathrm{ini}, y}=0 \text{ }\text{ } (p_y), \\

\sum\limits^N_{i=1} p_{i}\cos\theta_{i}-p_{\mathrm{ini}, z}=0 \text{ }\text{ } (p_z), \\

\sum\limits^N_{i=1} \sqrt{p_{i}^2+m_{i}^2}-E_{\mathrm{ini}}=0 \text{ }\text{ } (E).
\end{cases}
\end{equation}

\subsubsection{3C: Four-momentum conservation in a displaced vertex}
\label{sec:3C}

The 3C constraint utilizes four-momentum conservation at a given decay vertex, where a mother particle (M) decays weakly into $N$ particles. This procedure relies on measured information about the primary vertex and the decay vertex, and on a mass hypothesis of the mother particle as \textit{e.g.} $\Lambda$ or $K_s$. The angles of the mother particle, $\theta_M$ and $\phi_M$, are determined by the direction of the vector pointing from the primary to the decay vertex and are therefore treated as measured parameters, whereas the momentum $p_M$ is unmeasured and hence obtained from the fit. This results in three over-constraints (3C). The initial value is estimated from energy conservation of the decay products using

\begin{equation}
    p_{M}=\sqrt{\qty(\sum_i\sqrt{p_{i}^2+m_{i}^2})^2-m^2_{M}}\; .
    \label{eq:lambda_momentum}
\end{equation}
\ \\
In this expression, the subscript $M$ represents the mother particle, while $i$ refers to the decay products. The user must provide the mass of the mother particle as a mass hypothesis. The uncertainties in the vertex positions must estimated in the previous step and are propagated to the uncertainties in the parameters.  Since the momentum of the mother particle $p_M$ is an unmeasured quantity, its uncertainties are not known a priori but are, as the momentum itself, obtained from the fit. The constraint equations ensure four-momentum conservation in the decay of the mother:

\begin{equation}
f =
\begin{cases}

\sum\limits^N_{i=1} p_{i}\sin\theta_{i}\cos\phi_{i}-p_{M}\sin\theta_{M}\cos\phi_{M}=0 \text{ }\text{ } (p_x), \\

\sum\limits\limits^N_{i=1} p_{i}\sin\theta_{i}\sin\phi_{i}-p_{M}\sin\theta_{M}\sin\varphi_{M}=0 \text{ }\text{ } (p_y), \\

\sum\limits^N_{i=1} p_{i}\cos\theta_{i}-p_{M}\cos\theta_{M}=0 \text{ }\text{ } (p_z), \\

\sum\limits^N_{i=1} \sqrt{p_{i}^2+m_{i}^2}-\sqrt{p_{M}^2+m_{M}^2}=0 \text{ }\text{ } (E).

\end{cases}
\label{eq:3C}
\end{equation}

\subsubsection{1C: Four-momentum conservation with a missing particle}
\label{sec:missing}
Four-momentum conservation at the interaction point enables the reconstruction of one undetected or missing (indicated by the suffix \texttt{miss}) particle with a mass hypothesis $m_\mathrm{miss}$. I all other inital and final state particles are known or measured, the four-momentum conservation results in the following constraint equations:

\begin{equation}
f =
\begin{cases}

\sum\limits^N_{i=1} p_{i}\sin\theta_{i}\cos\phi_{i} + p_{\mathrm{miss},x}-p_{\mathrm{ini}, x}=0 \text{ }\text{ } (p_x), \\

\sum\limits^N_{i=1} p_{i}\sin\theta_{i}\sin\phi_{i} + p_{\mathrm{miss},y}-p_{\mathrm{ini}, y}=0 \text{ }\text{ } (p_y), \\

\sum\limits^N_{i=1} p_{i}\cos\theta_{i} + p_{\mathrm{miss},z}-p_{\mathrm{ini}, z}=0 \text{ }\text{ } (p_z), \\

\sum\limits^N_{i=1} \sqrt{p_{i}^2+m_{i}^2} + \sqrt{p_\mathrm{miss}^2+m_\mathrm{miss}^2}-E_{\mathrm{ini}} \\ \hspace{20pt}=0 \text{ }\text{ } (E).
\end{cases}
\end{equation}
\ \\
Here one has measured particles with indices $i$, a missing particle with mass $m_\mathrm{miss}$ and the initial beam-target four vector $p^\mu_\mathrm{ini}$. The mass, $m_\mathrm{miss}$ needs to be provided by the user as a hypothesis as well as $p^\mu_\mathrm{ini}$.

The four-momentum of the missing particle can be extracted after the fit. Since there are three unmeasured variables, \textit{i.e.} the three-momentum of the missing particle, only one over-constraint (1C) remains.

\subsubsection{1C: Missing Mass Constraint}
\label{sec:missingmass}

The missing mass constraint is suitable in the case when there is one undetected final state particle whose momentum is not of interest. Instead, a missing mass constraint can be formulated from the mass hypothesis of the missing particle: 

\setfirstterm{f}
\begin{multline}
\firstterm\:=\: \sqrt{\qty(E_\mathrm{ini}-\sum^N_{i=1}{E_{i}})^2 - \qty(\vec{p}_\mathrm{ini}-\sum^N_{i=1}{\vec{p}_{i}})^2}\\ 
\ftspace\:- m_\mathrm{miss} = 0.
\end{multline}

\subsubsection{1C: Invariant Mass Constraint}
\label{sec:mass}

The mass constraint can be used to constrain a set of particles to originate from a common mother particle, whose mass is known. The invariant mass of this set of particles is then constrained to the mass of the hypothetical mother particle:

\begin{equation}
    f = \sqrt{\qty(\sum^N_{i=1}{E_{i}})^2 - \qty(\sum^N_{i=1}{\vec{p}_{i}})^2} - m_\mathrm{mother} = 0 \; .
\end{equation}

\subsection{Convergence} \label{subsec:convergence}

The iterative fitting procedure is terminated when convergence is achieved. There are three different convergence criteria: The difference in $\chi^2$ of the Lagrange function between consecutive iterations, the Euclidean norm of the sum of all constraint equations and the Euclidean norm of the difference of all track parameters, normalized to the initial measurement, between two consecutive iterations. The default value for the convergence criteria is $10^{-4}$ but can be customized. If convergence is not reached before a specified number of iterations (default 20), the fitting procedure is exited.

\subsection{Goodness of fit}
The performance of the fit is assessed by the final $\chi^2_{final}$ of the fit and the so-called pull distributions for all fitted variables. The value of $\chi^2_{final}$ should be small, on the order of the number of over-constraints, $N_C$. For an ensemble of events, $N_C$ defines the shape of the $\chi^2_{final}$ distribution. There is a one-to-one relation between $\chi^2_{final}$ for a given $N_C$ and the corresponding probability, as defined by the probability density function \cite{PDG}. Ideally, the probability distribution should be uniform if the correct particle hypotheses have been used in the fit and if the covariance matrices accurately describe the measurement precision. However, if the particle hypotheses are incorrect, then a peak towards zero probability should be discernible. If elements of the covariance matrix are over- or underestimated, the distribution will decrease or increase towards larger values of probability. 

The pull distributions are defined as:

\begin{equation}
pull=\frac{\eta-y}{\sqrt{\sigma^2(y)-\sigma^2(\eta)}} \; ,
\end{equation}
for each variable, where $\eta$/$y$ are the fitted/measured variables, respectively, and $\sigma(\eta)$/$\sigma(y)$ are the corresponding uncertainties. The pull distribution should follow a normal distribution with a mean value of 0 and a standard deviation of 1.

\section{Class Descriptions} \label{sec:classes}
The \lstinline|KinFit| package is written in C++ and based on ROOT~\cite{root} (version 6) and uses CMake~\cite{cmake} (version 3.0 or newer) for the installation. It is available online at:
\begin{lstlisting}
https://github.com/KinFit/KinFit.git
\end{lstlisting}
Documentation about the usage of the provided tools can be found in the README of the git repository and in the User's Guide in~\Cref{app:usersguide}.\

The properties of the particle candidates needed as input are the particle's track parameters, mass, and covariance matrix. Functions are provided to transform Cartesian track parameters to the $R, Z$ track parameters.

\subsection*{KFitParticle}
The track parameters of particle candidates are organized in \lstinline|KFitParticle| objects that provide all information needed by the fitter class. It inherits from the ROOT class \lstinline|TLorentzVector|. For each candidate, the values for the track parameters described in Section~\ref{sec:trackparametrization} have to be set. Optionally, an arbitrary particle ID and track ID can be chosen by the user for later reference.

\subsection*{KFitDecayCandFinder}

\lstinline|KFitDecayCandFinder| calculates properties of an unmeasured candidate that decays in a displaced vertex to be used later in the fit. The angles, $\theta$ and $\phi$ are calculated from the line segment connecting the primary to the displaced decay vertex. The uncertainties for these angles are propagated from the uncertainties in the vertex positions in X, Y and Z using the matrix formalism for error propagation. The user needs to provide the uncertainties for the vertex positions.

\subsection*{KFitVertexFinder}

\lstinline|KFitVertexFinder| finds the vertex by calculating the point of closest approach between at least two tracks by a matrix formalism. For more than two tracks, the vertex is taken as the center of gravity, \textit{i.e.} the point that is simultaneously closest to all tracks.. This is calculated from a least square method. The vertex finding code is based on a procedure in HYDRA \footnote{https://subversion.gsi.de/hades/hydra2/}, the HADES software.

\subsection*{KinFitter}

\lstinline|KinFitter| is the main class containing the fitting functions. It can perform a vertex, 4C, 3C, missing particle, missing mass, mass and a combined vertex+mass fit (see Section~\ref{sec:constraints}). The constraint equations and Jacobi matrices are implemented here. This is where the iterative fitting procedure is carried out. The maximum number of iterations or convergence criteria can be changed to custom values.

\subsection*{KFitAnalyzer}

\lstinline|KFitAnalyzer| is a user interface class. It contains user settings and performs the event loop with a fit of choice. The input particles need to be stored as \lstinline|KFitParticles| in a \lstinline|TClonesArray|. For each event, particles are selected based on their particle ID (PID), which is also provided to the \lstinline|KFitAnalyzer|. A \lstinline|KFitDecayBuilder| object is created which takes the selected particles and the desired constraint as input. In the end the fitted particles are retrieved from the \lstinline|KFitDecayBuilder| and stored in an output file together with the fit probability. The \lstinline|KFitAnalyzer| currently performs all fits except the 3C fit.

\subsection*{KFitDecayBuilder}

This class is responsible for building all possible combinations of the particles within one event. Each combination is passed to the \lstinline|KinFitter| which performs the selected fit. The \lstinline|KFitDecayBuilder| selects the combination with the best fit probability.

\section{Performance Study} \label{sec:performance}
The performance of \lstinline|KinFit| is benchmarked using Monte Carlo (MC) simulations. These represent an ideal scenario where a finite detector resolution is added by smearing, and thus known, whereas in real data, it might be difficult to estimate the covariance matrix exactly. This approach allows for a focused investigation of the quality of the \lstinline|KinFit| tools. The reaction investigated is $pp\rightarrow pK^+\Lambda$, $\Lambda\rightarrow p\pi^-$, as illustrated in Figure~\ref{fig:lambda_vertices}. This reaction provides an opportunity to examine all tools contained in the package. All fits performed in this section use the default settings of \lstinline|KinFitter|, described in Section~\ref{subsec:convergence}.\\

The event generator Pluto~\cite{pluto} was used to generate $100\,000$ events of the $pp\rightarrow pK^+\Lambda$ $\Lambda\rightarrow p\pi^-$ reaction at a beam kinetic energy of T=$4.5\,\mathrm{GeV}$. In our simulations, the produced particles and their decay products are distributed isotropically across the available phase space. The primary vertex is generated at the origin, \textit{i.e.} (0,0,0). A full 4$\pi$ acceptance is assumed and no material effects are included. The $\Lambda$ hyperons decay at a displaced decay vertex according to the mean $\Lambda$ hyperon life time $c_\tau(\Lambda) = 7.98\,\mathrm{cm}$~\cite{PDG}. The track parameters of the final state particles are smeared according to a Gaussian distribution, simulating uncertainties from \textit{e.g.}, detector resolution in a controlled way. The uncertainty for each track parameter is listed in Table~\ref{tab:smearing}. Since the results presented here do not depend on a specific experiment, the uncertainties are set to be constant, except the momentum uncertainty, which is momentum dependent.\\

\begin{table*}[ht]
    \centering
    \begin{tabular}{l|c|c|c|c|c}
        Track parameter &  $1/p$ & $\theta$ & $\phi$ & $R$ & $Z$ \\
        \hline
        $\sigma$ & $0.025\cdot 1/p$ & $0.0009\,\mathrm{rad}$ & $0.0009\,\mathrm{rad}$ & $0.5\,\mathrm{mm}$ & $1\,\mathrm{mm}$ 
    \end{tabular}
    \caption{Gaussian standard deviations of the track parameters in the MC sample after smearing.}
    \label{tab:smearing}
\end{table*}

\subsection{Mass Fit using KFitAnalyzer}\label{sec:mass_fit}
\begin{figure}
    \centering
    \includegraphics[width= 0.5\textwidth]{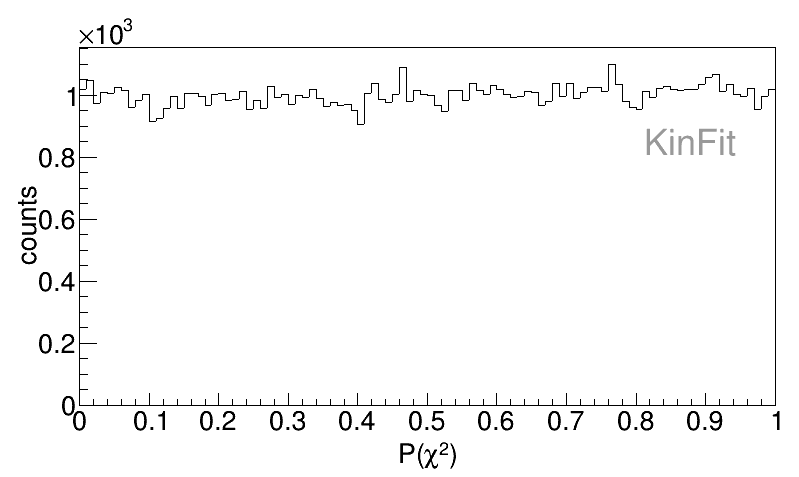}
    \includegraphics[width= 0.5\textwidth]{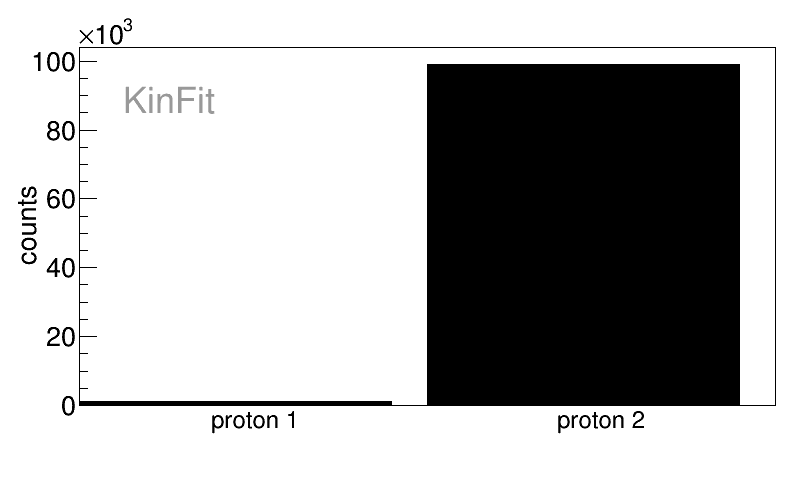}
    \caption{Probability distribution of a $\Lambda$ hyperon mass fit in the reaction $pp\rightarrow pK^+\Lambda$ $\Lambda\rightarrow p\pi^-$ using \lstinline|KFitAnalyzer| (top) and the proton selected to originate from the $\Lambda$ hyperon decay (bottom)}
    \label{fig:mass_fit}
\end{figure}

In each event, there are four measured particles: a proton ($p_1$) and a kaon from the interaction point and a pion and a proton ($p_2$) from the $\Lambda$ hyperon decay. This means there are two possibilities to combine a pion with a proton, either $p_1\pi^-$ and $p_2\pi^-$, of which the latter is the correct $\Lambda$ decay candidate. The \lstinline|KFitAnalyzer| is used to find the pion-proton combination that comes from the $\Lambda$ hyperon decay through the application of a mass fit.\\
The pion-proton pair giving the largest mass fit probability (see Fig.~\ref{fig:mass_fit}) are selected as $\Lambda$ daughters. In $99\,\%$ of the events, the correct proton $p_2$ is chosen. The probability distribution is uniform, as expected.

\subsection{Reconstruction of  \texorpdfstring{$\Lambda$}{lg} Hyperon from Vertex Positions and a 3C Fit}
The $\Lambda$ hyperon reconstruction procedure includes three steps:
\begin{enumerate}
    \item \textbf{Vertex finding}\\
        As a first step the position of the primary vertex, where the $\Lambda$ hyperon was produced is determined from the point of closest approach between the other tracks coming from the production vertex. In this case, these are the proton ($p_1$) and kaon tracks. This task is performed by \lstinline|KFitVertexFinder|. In a similar way, the $\Lambda$ hyperon decay vertex is obtained from the  point of closest approach between the proton ($p_2$) and pion tracks originating from the $\Lambda$ hyperon decay. The vertex positions are shown in Fig.~\ref{fig:vertices}. All primary particles are produced at the origin which means that the distribution in the top panel reflects the finite resolution of the vertex estimation. The dominating effect in the $\Lambda$ decay vertex distribution is the distance travelled by $\Lambda$ before decaying. 
        \begin{figure}
            \centering
            \includegraphics[width= 0.5\textwidth]{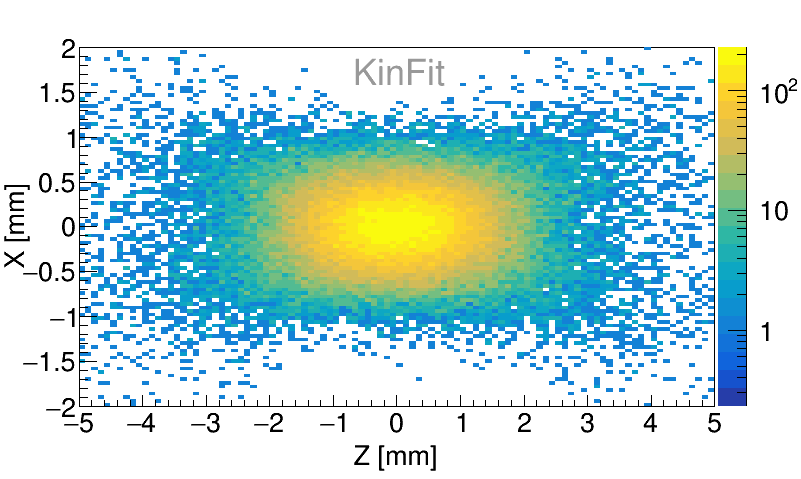}
            \includegraphics[width= 0.5\textwidth]{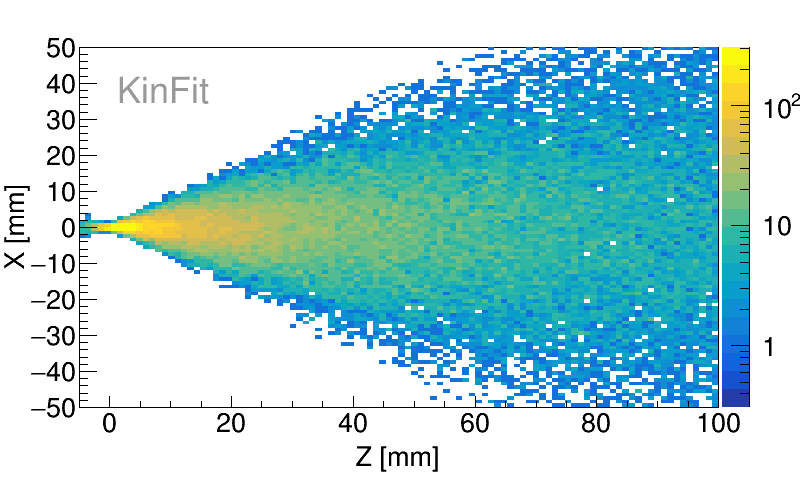}
            \caption{Reconstructed production vertex from proton and $K^+$ (top) and location of reconstructed decay vertices of the $\Lambda$ hyperon from the secondary proton and $\pi^-$ (bottom) as reconstructed by the \lstinline|KFitVertexFinder|. The positions are projected onto the x-z plane.}
            \label{fig:vertices}
        \end{figure}
    \item \textbf{Reconstruction of the neutral candidate}\\
        The direction of the $\Lambda$ candidate is given by the vector pointing from the primary to the decay vertex. The magnitude of the $\Lambda$ hyperon momentum is estimated by the  of the sum of the  three-momenta of its daughter particles, see \Cref{eq:lambda_momentum}. The \lstinline|KFitDecayCandFinder| carries out this task and calculates the uncertainties of the track parameters of the $\Lambda$ candidate. The latter requires information on the vertex uncertainties which were estimated by Gaussian fits to the vertex resolution in $x$, $y$ and $z$ for each vertex.
    \item \textbf{3C fit}\\
        A kinematic fit ensuring four-momentum conservation at the $\Lambda$ hyperon decay vertex is the final step and is executed by \lstinline|KinFitter|. This improves the resolution of the $\Lambda$ hyperon track parameters significantly (Figure~\ref{fig:resolutions}) and can be used to reject false combinations of protons in the vertices.
        \begin{figure*}
            \centering
            \includegraphics[width=1.0\textwidth]{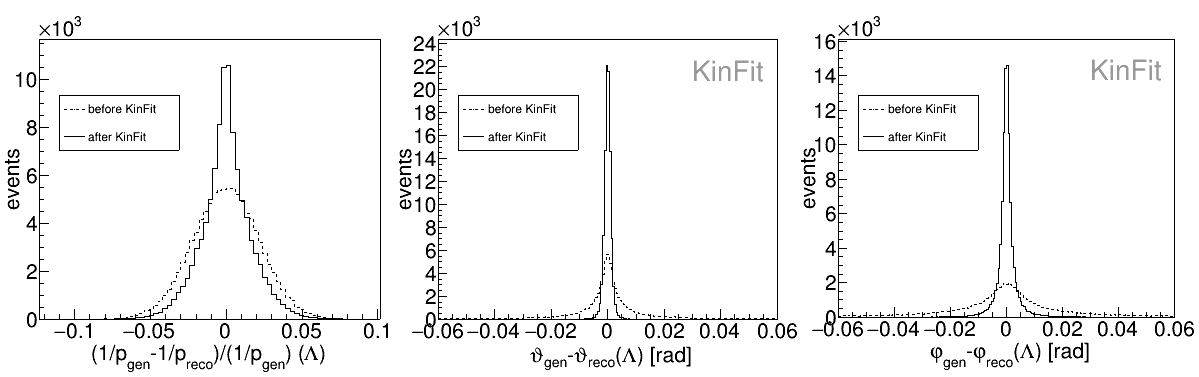}
            \caption{Resolution of the $\Lambda$ hyperon track parameters before and after the 3C fit.}
            \label{fig:resolutions}
        \end{figure*}
\end{enumerate}

\begin{figure*}
    \centering
    \includegraphics[width= 1.0\textwidth]{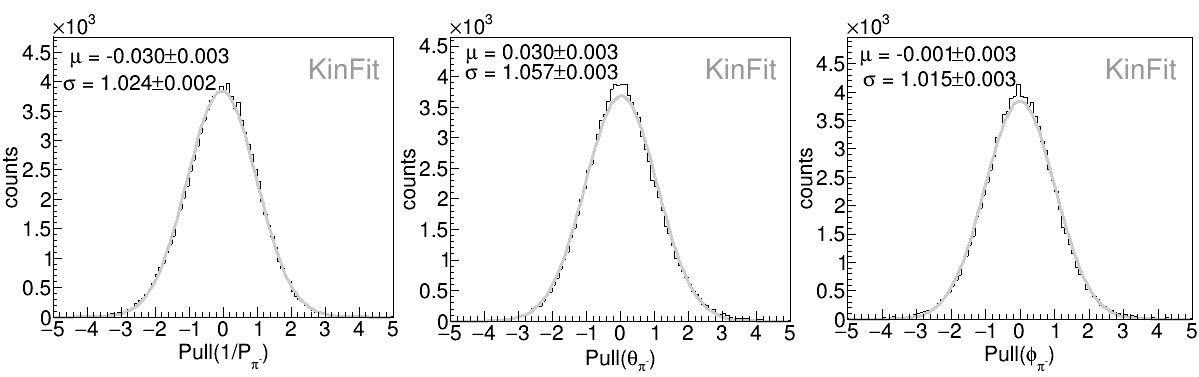}
    \caption{Pull distributions for the $\pi^-$ track parameters after the 3C fit with respective mean and standard deviation.}
    \label{fig:pulls}
\end{figure*}

\begin{figure}
    \centering
    \includegraphics[width= 0.5\textwidth]{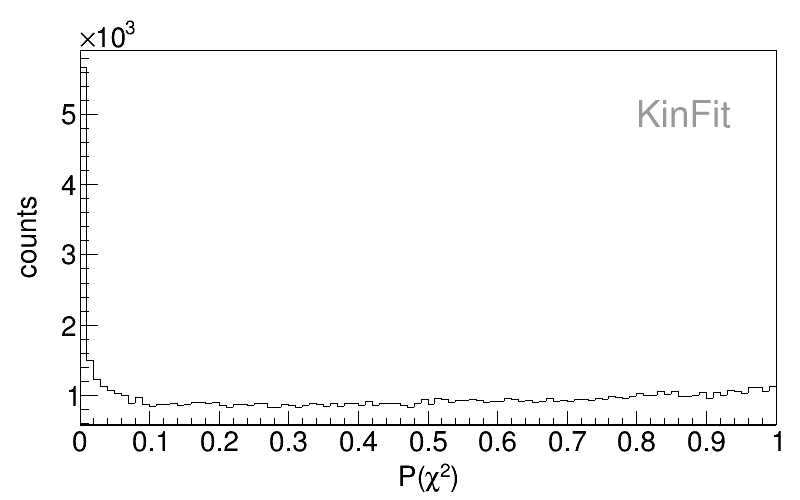}
    \caption{Probability distribution from the 3C fit.}
    \label{fig:prob}
\end{figure}

Only the correct combination of particles was used in this example, i.e. correct assignment of the protons to the vertices. The pull distributions for the pion track are shown as an example in Fig.~\ref{fig:pulls}. These nearly follow a normal distribution, as expected. Fig.~\ref{fig:prob} shows the probability distribution of the 3C fit. The probability deviates slightly from a uniform distribution and there are some events for which the fit converged, but gives a low probability. This effect and the slightly larger deviation of the pull distributions from a normal distribution are only seen in the 3C fit and originate from the estimation of the vertex resolutions, which are not exactly described by a Gaussian.

\subsection{4C Fit of the Reaction  \texorpdfstring{$pp\rightarrow pK^+\Lambda$}{lg}}
\label{sec:4C_test}
\begin{figure}
    \centering
    \includegraphics[width= 0.5\textwidth]{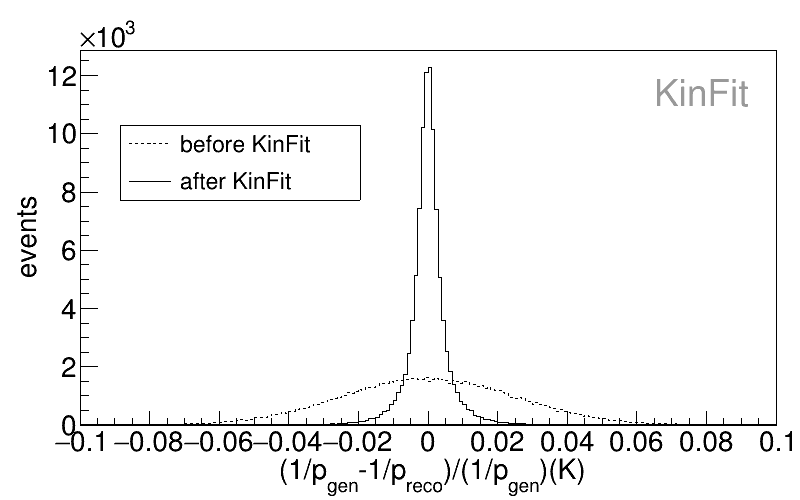}
    \includegraphics[width= 0.5\textwidth]{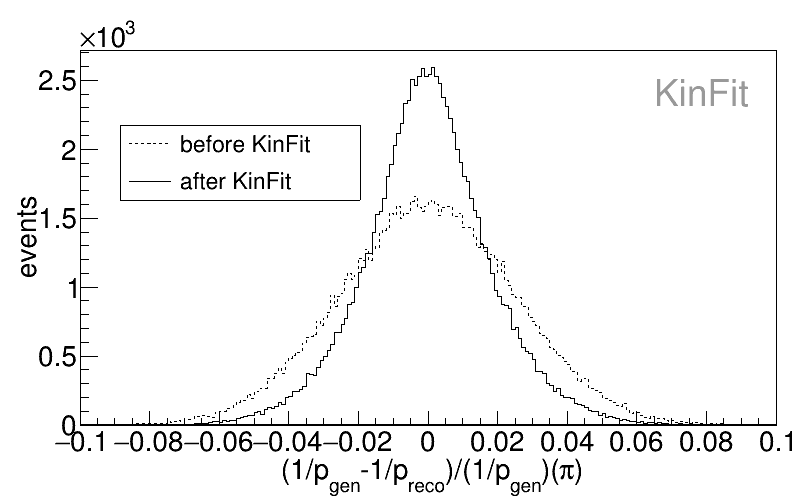}
    \caption{Momentum resolution for the $K^+$ (top) and $\pi^-$ (bottom) before and after the 4C fit.}
    \label{fig:res_4c}
\end{figure}

\begin{figure}
    \centering
    \includegraphics[width= 0.5\textwidth]{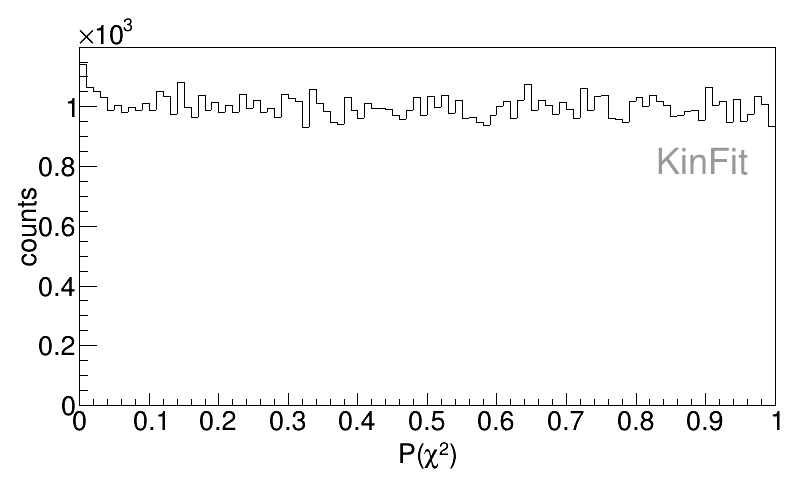}
    \includegraphics[width= 0.5\textwidth]{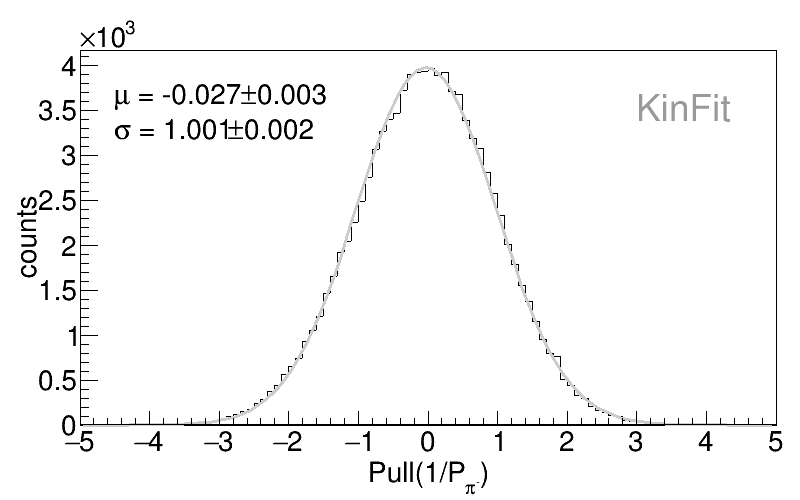}
    \caption{4C fit. Probability distribution (top) and example pull distribution of the $\pi^-$ momentum with mean and standard deviation (bottom).}
    \label{fig:quality_4c}
\end{figure}

The 4C fit is used to constrain the four-momenta of all final state particles to that of the initial beam-target system. In this example, the final state particles are a proton and a kaon from the primary vertex and a proton and a pion from the $\Lambda$ hyperon decay vertex. Fig.~\ref{fig:res_4c} shows the momentum resolution for the kaon and the pion before and after the fit. The improvement is more substantial for the kaon momentum resolution compared to the other particles. This is due to the larger uncertainty of the kaon associated with its larger total momentum. The probability distribution is uniform and the pull distributions follow a normal distribution, as shown in Figure~\ref{fig:quality_4c}.

\subsection{Missing Particle Fit of  \texorpdfstring{$K^+$}{lg} in the Reaction  \texorpdfstring{$pp\rightarrow pK^+\Lambda$}{lg}.}
\begin{figure}
    \centering
    \includegraphics[width= 0.5\textwidth]{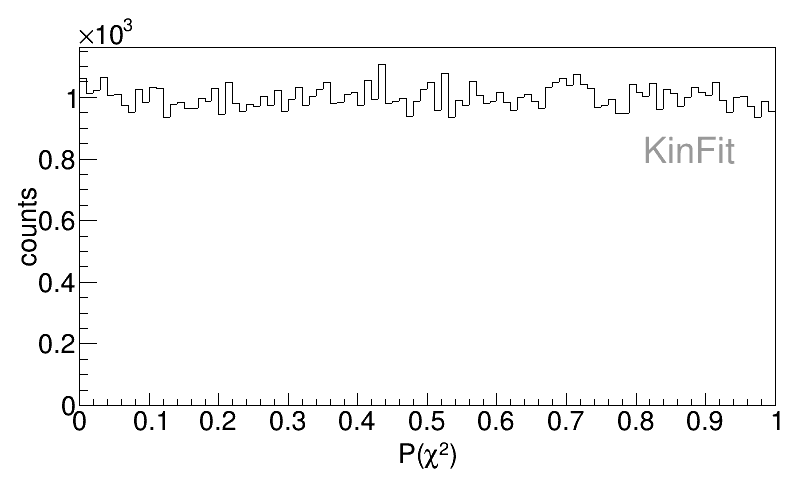}
    \includegraphics[width= 0.5\textwidth]{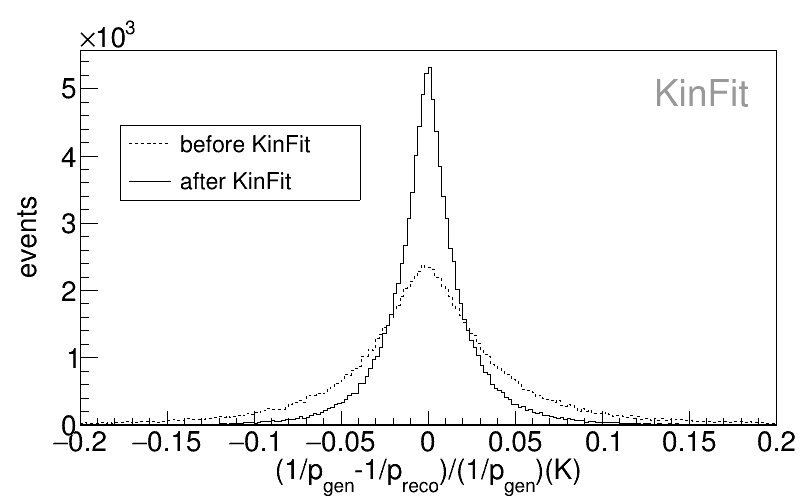}
    \caption{Missing particle fit. Probability distribution (top) and resolution of the estimated and fitted $K^+$ momentum (bottom).}
    \label{fig:quality_miss}
\end{figure}

In this scenario, it is assumed that the kaon is not detected. The missing particle fit is employed to constrain the four-momenta of the detected particles along with the undetected particle to match the beam-target system, as well as to estimate the momentum of the missing particle. Figure~\ref{fig:quality_miss} presents the probability distribution of the fit and compares the kaon momentum resolution from the initial guess and after the fit. The initial guess for the kaon momentum is calculated from three-momentum conservation in the interaction point. The momentum resolution before the fit is worse than in the example in~\Cref{sec:4C_test}, where the momentum was measured directly. A noticeable improvement in the momentum resolution can be observed after the fit. However, the resolution is still worse than in the case of the 4C fit, as the missing particle fit has fewer over-constraints. The resolutions of the other track parameters and the pull distributions appear quite similar to those of the 4C fit.

\subsection{Vertex Fit}
\begin{figure}
    \centering
    \includegraphics[width= 0.5\textwidth]{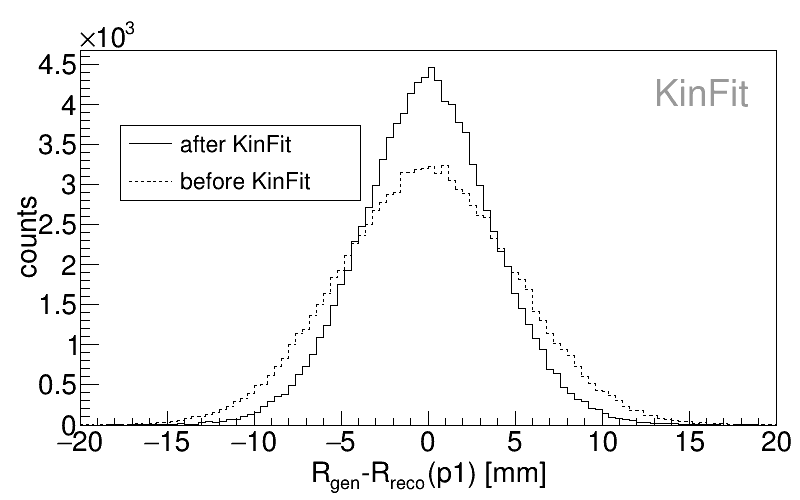}
    \caption{Vertex fit. Resolution of the R-parameter of the primary proton ($p_1$) before and after the fit.}
    \label{fig:quality_vtx_R_zvtx}
\end{figure}

\begin{figure}
    \centering
    \includegraphics[width= 0.5\textwidth]{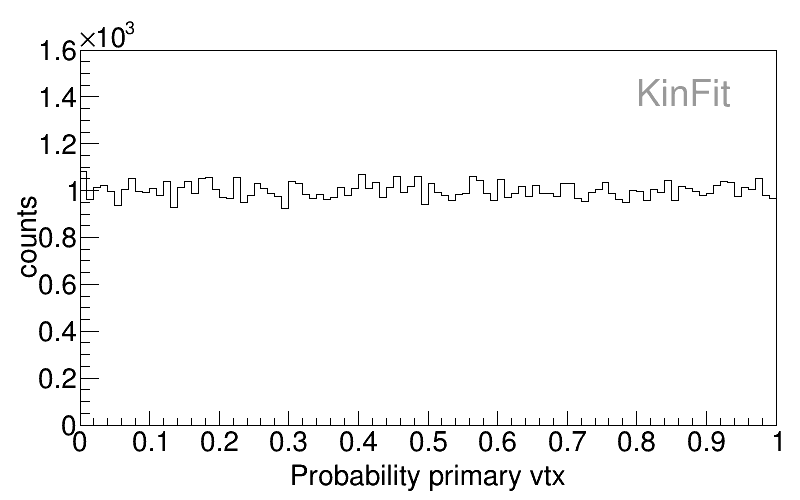}
    \includegraphics[width= 0.5\textwidth]{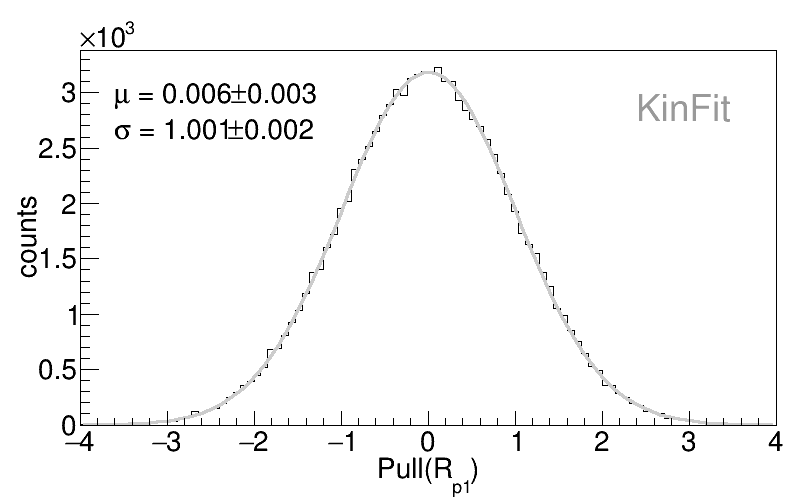}
    \caption{Vertex fit. Probability distribution (top) and example pull distribution of the R-parameter for the primary proton ($p1$) (bottom).}
    \label{fig:quality_vtx_prob_R}
\end{figure}

The vertex fit aims to constrain the track parameters of final state particles that originate from the same single point in space. To perform the vertex fit, at least two outgoing particles from the same vertex must be measured. In this example, these particles are the kaon and proton ($p_1$) produced at the beam-target interaction point. Fig.~\ref{fig:quality_vtx_R_zvtx} illustrates the resolution of the estimated R-parameter, defined in Section \ref{sec:trackparametrization}, for the proton. The probability distribution for this vertex fit, along with an example pull distribution of the R-parameter of $p_1$, is depicted in Fig.~\ref{fig:quality_vtx_prob_R}. As anticipated, the probability is uniformly distributed across its range, while the pull follows a normal distribution. The \lstinline|KinFitAnalyzer| was used to choose the proton track that has the largest probability to come from the same vertex as the kaon track. The correct proton ($p_1$) is chosen in $85\,\%$ of the events, which is lower than the fraction of correctly identified combinations by the mass fit presented in \Cref{sec:mass_fit}. This shows that the mass constraint is more powerful than the vertex fit.

\subsection{Runtime Performance}
The runtime performance of various components of \lstinline|KinFit| was evaluated on a laptop with the following hardware specifications:
\begin{itemize}
    \item Processor: Intel Core i7-1185G7, 3.00 GHz
    \item Memory: 32 GB RAM
\end{itemize}
The results are displayed in Table~\ref{tab:runtime}. The average runtime per 1000 events was calculated for those in which the fit converged. A low overhead of approximately a microsecond was observed.

\begin{table*}[h!]
    \centering
    \resizebox{\textwidth}{!}{\begin{tabular}{l|c|c|c|c|c}
        Stage &  Fitting & Initialization & Vertex Finding & Mother Candidate Finding \\
        \hline
        Time / event [$\mu$s] & 10 & 1 & 1 & 1
    \end{tabular}}
    \caption{Runtime per event for different parts of \lstinline|KinFit|.}
    \label{tab:runtime}
\end{table*}

\section{Conclusions}
The \lstinline|KinFit| package is a versatile kinematic fitting package that has been developed to provide tools for reconstruction of vertices, momenta and masses of particles in a generic hadron physics experiment. In particular, it is capable of fitting an unknown production vertex from an extended beam-target interaction volume, and of combining kinematic and geometric constraints. The tools provided by \lstinline|KinFit| have been tested with toy MC simulations, demonstrating outstanding performance in terms of improved track parameter resolutions and the ability to select correct particle hypotheses and -combinations. \\

While the focus has been on providing suitable tools for hadronic interactions and specifically hyperon decays, the package can be applied for other types of reactions as well. Though \lstinline|KinFit| uses the same track parametrization as HADES, but it is generally considered experiment-independent. \\

The minimal overhead introduced by adding \lstinline|KinFit| to an analysis suggests that the package is suitable for running on a local machine.

\section{Outlook}
Although the majority of particle physics analyses are based on \lstinline|C++|, the demand for similar \lstinline|Python| tools is currently growing due to the rapid expansion of the \lstinline|Python| ecosystem. Therefore, the \lstinline|KinFit| package is planned to be integrated into the SciKit-HEP project~\cite{Rodrigues:2020syo}, which is a community-driven \lstinline|Python| ecosystem for data analysis.

\section*{Author Contributions}
This work is conceptualized by Waleed Esmail and Karin Sch\"onning. The development and implementation of the C++ \lstinline|KinFit| package was done by Esmail, Jana Rieger and Jenny Taylor. The first paper draft preparation, was done by Esmail, Rieger and Taylor and the drafting was coordinated by Rieger. The project was supervised by Sch\"onning. Benchmark studies and bug fixes were performed by Rieger, Taylor and Malin Bohman, while all authors contributed to the review of the paper's intellectual content and the editing.

\section*{Acknowledgements}
We would like to thank Tord Johansson for useful discussions on kinematic fitting of $\Lambda$ hyperons. Furthermore, we are grateful to Anar Rustamonov for the suggestion to make KinFit available to a larger community as an independent package. Finally, we would like to express our appreciation of the fruitful discussions and the encouragements we received from James Ritman and Piotr Salabura.\\

The publication is funded by the Deutsche Forschungsgemeinschaft (DFG, German Research Foundation) – 491382106 , and by the Open Access Publishing Fund of GSI Helmholtzzentrum fuer Schwerionenforschung. This project has received funding from the Knut and Alice Wallenberg Foundation, Contract No. 2016.0157 and 2021.0299 (Sweden), and the Swedish Research Council, Contract No. 2019-04594 (Sweden).

\bibliography{cas-refs}

\newpage

\onecolumn
\appendix

\section{User's Guide} \label{app:usersguide}
The source code is available online in the \lstinline|KinFit| git repository. It can be downloaded via
\begin{lstlisting}
git clone https://github.com/KinFit/KinFit.git
\end{lstlisting} 
and installed using CMake. A ROOT 6 installation is required.\\

There are two options how the tools provided by this package may be applied by the user. One of them is to use the provided user functions through \lstinline|KFitAnalyzer|. These make it straight-forward to apply the fitter using basic constraints. However since the type of analysis where kinematic fitting can yield large improvements often requires a careful and customized event selection, the provided classes can also be applied directly by the user, which makes a fine tuning of the fitting process possible but requires a deeper understanding of the procedure.

\subsection{Automated application of a kinematic fit}
In order to use \lstinline|KinFit| in an automated way, the user needs to provide a root file with a \lstinline|TTree| called "data" that contains a \lstinline|TClonesArray| "KFitParticle" of \lstinline|KFitParticles| as input. The macro \lstinline|analysis_user.C| illustrates how to set up the \lstinline|KFitAnalyzer|.

{\begin{small}
 \begin{lstlisting}
#include "KFitAnalyzer.h"

Int_t analysis_user(TString infile, TString outfile, Int_t evts){

    KFitAnalyzer RootAnalyzer(infile, outfile, evts);
    std::vector<int> pids;
    pids.push_back(14); pids.push_back(9);
    Double_t mass = 1.11568;

    RootAnalyzer.doFitterTask("Mass", pids, mass);
    
    return 0;
}
\end{lstlisting}   
\end{small}}

A \lstinline|KFitAnalyzer| object is created. It takes the input file(s), the output file name and the number of events to be processed as input. The only command that needs to be executed to start the analysis is \lstinline|doFitterTask|. Here the type of fit, the PIDs of the particles to be fit and additional information that is required for the specific fit is requested. In this example, an invariant mass fit of a proton and a pion is performed, assuming that they originate from a $\Lambda$ decay. After the analysis procedure the fitted tracks of the best particle combination and the fit probability are written to the output file. This requires that the kinematic fit converges and that that combination yield the highest fit probability of the event. \\

Other available fits are
{\begin{small}
 \begin{lstlisting}
doFitterTask("4C", pids, -1, ppSystem);
doFitterTask("MassVtx", pids, mLambda);  doFitterTask("MissingMass", pids, mass, ppSystem);
doFitterTask("MissingParticle", pids, mass, ppSystem); 
doFitterTask("Vertex", pids, -1);

\end{lstlisting}   
\end{small}}

\subsection{Using the individual classes} \label{sec:individualusageofclasses}

In an event-based analysis, the kinematic fitting tools are typically applied as one of the first analysis steps inside the "event loop", \textit{i.e.} while iterating through all events, selecting the most suitable particle candidates from each. \\

\textbf{Creating a \lstinline|KFitParticle|}\\
After a set of particle candidates are selected whose track parameters are to be fit, a \lstinline|KFitParticle| object is created for each candidate. In addition, the covariance matrix has to be assigned to each candidate. This is done either by calling the constructor \lstinline|KFitParticle(TLorentzVector cand, double R, double Z)| or \lstinline|KFitParticle(TLorentzVector cand, double X, double Y, double Z)| and setting the covariance matrix using the \lstinline|setCovariance()| function directly, or by using a suitable \lstinline|FillData| function, useful when the covariance was estimated and is not known for each particle individually. An example of how a \lstinline|FillData| function can look like to set the attributes of the \lstinline|KFitParticle| is shown below.

\begin{lstlisting}
void FillData(TLorentzVector *cand, KFitParticle *outcand, double R, double Z, double arr[], double mass, int trackID, int pid)
    
    /** Cov(5,5)Covariance matrix of the KFitParticle
    * Diagonal entries correspond to the covariances 
    * in the parameters in the following order
    * 
    * -----------------------------
    * | 1/p                       |
    * |     theta                 |
    * |            phi            |
    * |                   R       |
    * |                        Z  |
    * -----------------------------
    * 
    * Off diagonal elements corresponds to the
    * correlations between the parameters
    * arr[] can be adjusted acordingly to set the
    * off diagonal elements
    */
    
    TMatrixD cov(5, 5);
    cov(0, 0) = std::pow(arr[0], 2);
    cov(1, 1) = std::pow(arr[1], 2);
    cov(2, 2) = std::pow(arr[2], 2);
    cov(3, 3) = std::pow(arr[3], 2);
    cov(4, 4) = std::pow(arr[4], 2);
    

    outcand->SetXYZM(
        cand->P() * std::sin(cand->Theta()) * std::cos(cand->Phi()),
        cand->P() * std::sin(cand->Theta()) * std::sin(cand->Phi()),
        cand->P() * std::cos(cand->Theta()), mass);
    outcand->setThetaRad(cand->Theta());
    outcand->setPhiRad(cand->Phi());
    // outcand->setThetaDeg(cand->Theta()); // Depending on the unit of the angles from cand the functions setThetaRad(double val) and setThetaDeg(double val) can be used iterchangably
    // outcand->setPhiDeg(cand->Phi()); // Depending on the unit of the angles from cand the functions setPhiRad(double val) and setPhiDeg(double val) can be used iterchangably
    outcand->setR(R);
    outcand->setZ(Z);
    outcand->setCovariance(cov);
    
    //optional
    outcand->setTrackId(trackID);
    outcand->setPid(pid);

\end{lstlisting}

In this example, R and Z are given as an input by the user. Alternatively the creation point of the particle candidate can be given in cartesian coordinates which are internally converted to $R$ and $Z$. \lstinline|SetXYZM()| sets the attributes of the \lstinline|TLorentzVector| that the \lstinline|KFitParticle| inherits from whereas \lstinline|setThetaRad()|,  \lstinline|setPhiRad()|, \lstinline|setR()|, \lstinline|setZ()| and \lstinline|setCovariance()| operate directly on the \lstinline|KFitParticle| object. The covariance matrix needs to be provided by the user as well as a mass hypothesis. A PID and track-ID can be set optionally.\\



\textbf{Handling the \lstinline|KinFitter|}\\
To use fitting functions, the \lstinline|KFitParticle| objects that will be used need to be placed in a vector of the type \lstinline|std::vector| from the C++ STD library. This vector needs to be passed to the constructor of \lstinline|KinFitter|. How this is done is illustrated in the examples below. For most fit options, an arbitrary number of particles can be added. However, for the vertex fit, exactly two particles need to be added. Following this, the user must choose one of the constraints, either; the 3C, the 4C, mass, missing mass, the vertex or the missing particle. Adjusting the number of iterations as well as the convergence criteria is possible but optional; if not set, the default criteria are applied. Then the \lstinline|fit()| function is called. This performs the actual fitting and returns \textit{true} if the fit converged. Some information can be obtained independently of the individual fit function that is used, \textit{e.g.} \lstinline|getChi2()| returns the $\chi^2$ of the fit and \lstinline|getProb()| returns the corresponding probability. \lstinline|getPull(int val)| returns the pull of variable $v$ for a daughter particle $p$, $val = 5\cdot d + v$. The function \lstinline|isConverged()| returns a \lstinline|boolean|; \textit{true} if the fit has converged and \textit{false} otherwise, this can be used for event or sample selections. The initialization of \lstinline|KinFitter| settings that can be used before calling \lstinline|fit()| are summarized below. 

\begin{lstlisting}
KinFitter(const std::vector<KFitParticle> &cands);
setNumberOfIterations(int val); // default 20
setConvergenceCriteria(double val, double val2, double val3);
// default 1e-4 
setVerbosity(int val); //between 0 and 2, default 0
\end{lstlisting}

After the fitting procedure called by \lstinline|fit()|, the fit result can be assessed by the following functions. Updated track parameters are obtained by the \lstinline|KFitParticle| objects returned by the first function

\begin{lstlisting}
KFitParticle getDaughter(int val); // Fitted daughter particle #val
double getChi2(); // Returns chi2
double getProb(); // Returns fit probability
double getPull(int val); // Returns pull of variable val
//* The pulls are returned in the following order: 
* val = i + 0 : 1/ p of particle i
* val = i + 1 : theta of particle i
* val = i + 2 : phi of particle i
* val = i + 3 : R of particle i
* val = i + 4 : Z of particle i
* Where the particles appear in the same order they 
* were entered into the fit
*/
bool isConverged(); // Returns true if fit converged
int getIteration(); // Returns number of iterations until convergence or maximal number of iterations
\end{lstlisting}

An example of how to perform the vertex fit is performed is shown below.





\begin{lstlisting}
#include "KinFitter.h"

KinFitter fitter(cand_vector);
fitter.addVertexConstraint(); // Choose vertex constraint
fitter.fit();
\end{lstlisting}

To perform the missing particle fit, the mass of the missing particle and a \lstinline|TLorentzVector| corresponding to the initial beam-target system must be given as input to the fitter in addition to a mass hypothesis. An example of how the \lstinline|TLorentzVector| can be constructed is in the following way: 

\lstinline|TLorentzVector ppSystem(p1,p2,p3,E);|

where the first three entries correspond to the momentum in each Cartesian direction and the last entry is the total energy of the system. After the fit has been performed, the missing daughter can be retrieved as a \lstinline|TLorentzVector|

\begin{lstlisting}
#include "TLorentzVector.h"
#include "KinFitter.h"

Double_t mass;
TLorentzVector ppSystem(p1,p2,p3,E);
KinFitter fitter(cand_vector);
void addMissingParticleConstraint(ppSystem, mass); // Choose momentum constraint for missing particle
fitter.fit();
TLorentzVector getMissingDaughter(); // Retrieve the missing daughter after the fit
\end{lstlisting}

To perform the 4C fit, in addition to the vector of final state particles, a \lstinline|TLorentzVector| corresponding to the initial beam-target system must be passed to the fitter.

\begin{lstlisting}
#include "TLorentzVector.h"
#include "KinFitter.h"

TLorentzVector ppSystem(p1,p2,p3,E);
KinFitter fitter(cand_vector);
fitter.add4Constraint(ppSystem); // Choose 4C fit
fitter.fit();
\end{lstlisting}

To perform a mass fit, in addition to the vector of final state particles, the mass of the particle must be passed to the fitter.

\begin{lstlisting}
#include "KinFitter.h"

Double_t mass;
KinFitter fitter(cand_vector);
fitter.addMassConstraint(mass); // Choose mass fit
fitter.fit();
\end{lstlisting}

To perform a missing mass fit, in addition to the vector of final state particles, the mass of the missing particle and a \lstinline|TLorentzVector| corresponding to the initial beam-target system must be passed to the fitter.

\begin{lstlisting}
#include "TLorentzVector.h"
#include "KinFitter.h"

Double_t mass;
TLorentzVector ppSystem(p1,p2,p3,E);
KinFitter fitter(cand_vector);
fitter.addMissingMassConstraint(ppSystem, mass); // Choose missing mass fit
fitter.fit();
\end{lstlisting}

Since a mother particle needs to be constructed in order to perform the 3C fit, this is a bit more involved and an example of how to perform this fit and run the 3C fit is given below

\begin{lstlisting}
#include "KinFitter.h"
#include "KFitVertexFinder.h"
#include "KFitDecayCandFinder.h"

#include "TLorentzVector.h"

#include <vector>
#include <algorithm>
#include <map>
#include <iostream>
#include <iomanip>
#include <math.h>

// Create a vector for particle pairs from the interaction point and decay vertex, respectively
std::vector<KFitParticle> cands1, cands2;
cands1.push_back(proton1_fit);
cands1.push_back(kaon_fit);
cands2.push_back(proton2_fit);
cands2.push_back(pion_fit);

// Initialize a KFitVertexFinder for each vertex
KFitVertexFinder vtx1finder(cands1);
KFitVertexFinder vtx2finder(cands2);

// Find vertex and retrieve it
TVector3 vtx1 = vtx1finder.getVertex();
TVector3 vtx2 = vtx2finder.getVertex();

// Vertex resolutions
Double_t vtx1_xres, vtx1_yres, vtx1_zres, vtx2_xres, vtx2_yres, vtx2_zres; 

// Find the decaying candidate
KFitDecayCandFinder lambdafinder(cands2, 1.115683, vtx1, vtx2, vtx1_xres, vtx1_yres, vtx1_zres, vtx2_xres, vtx2_yres, vtx2_zres); 
KFitParticle lambda_cand = 
    lambdafinder.getDecayCand();

// Do 3C fit in decay vertex
KinFitter fitter(cands2);
fitter.add3Constraint(lambda_cand);
fitter.fit();

// Get fit result
KFitParticle fcand1 = fitter.getDaughter(0); 
    // proton
KFitParticle fcand2 = fitter.getDaughter(1); // pion
KFitParticle lambda_fit = fitter.getMother(); 
    // lambda
TMatrixD test = lambda_fit.getCovariance(); 
    // lambda covariance
\end{lstlisting}

\pagebreak
\section{QA Plots} \label{app:plots}
\subsection{Pull Distributions of 4C Fit}
\begin{figure}[!ht]
    \centering
    \includegraphics[width= 0.34\textwidth]{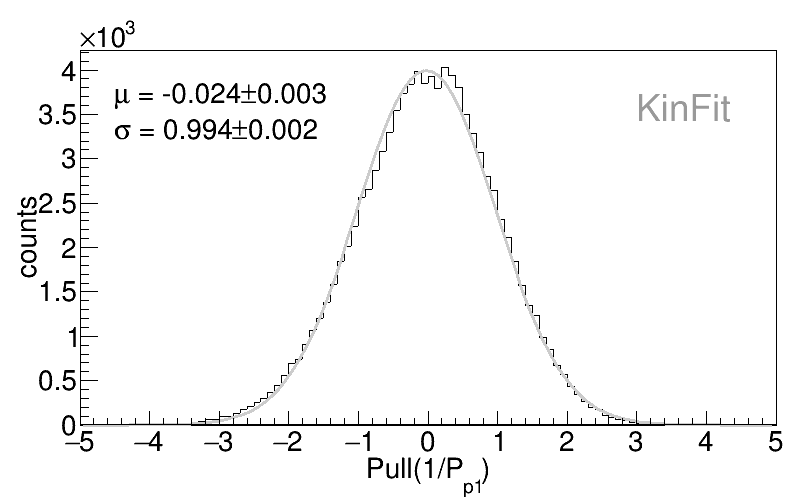}~
    \includegraphics[width= 0.34\textwidth]{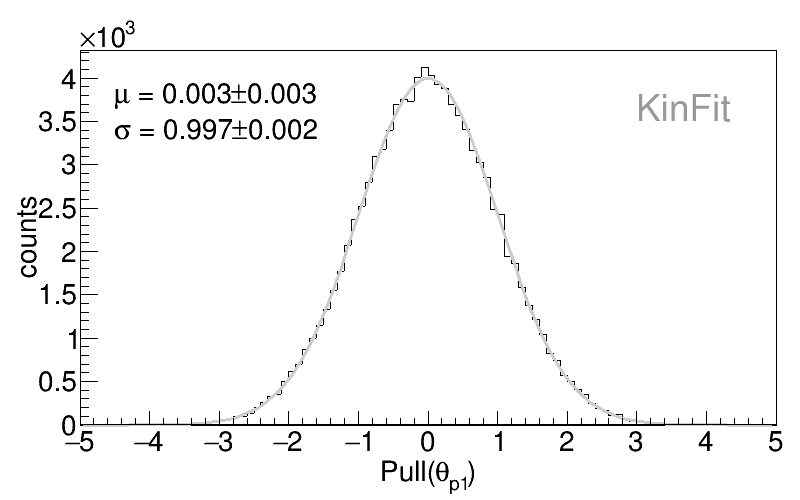}~
    \includegraphics[width= 0.34\textwidth]{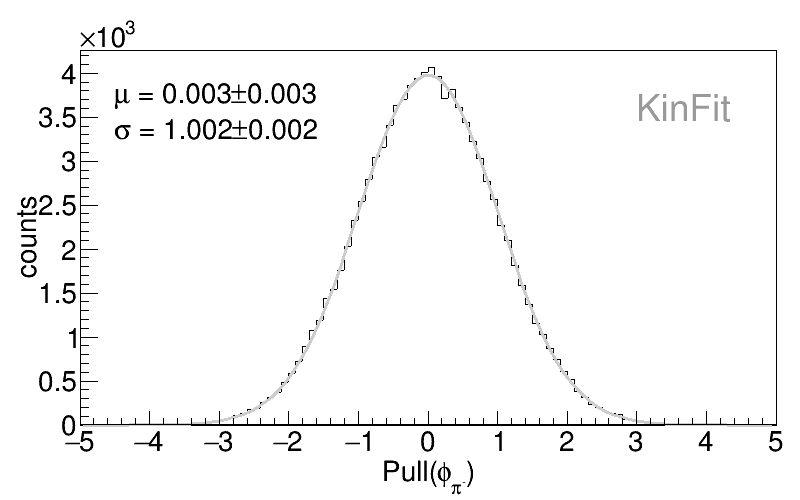}
    \includegraphics[width= 0.34\textwidth]{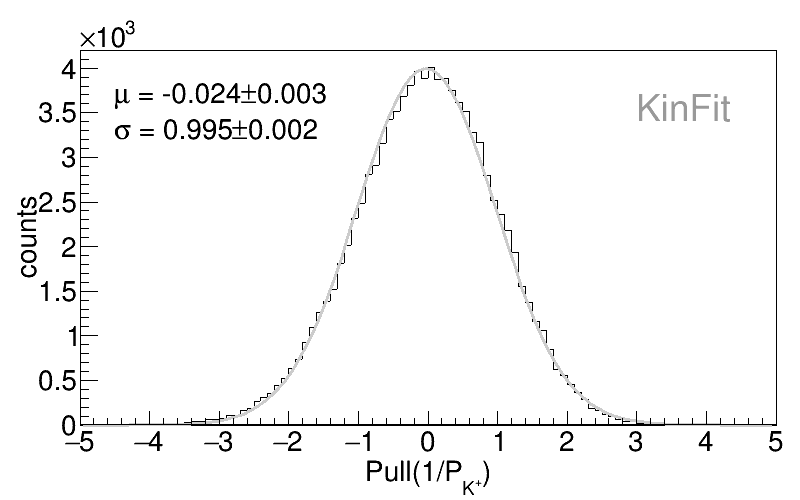}~
    \includegraphics[width= 0.34\textwidth]{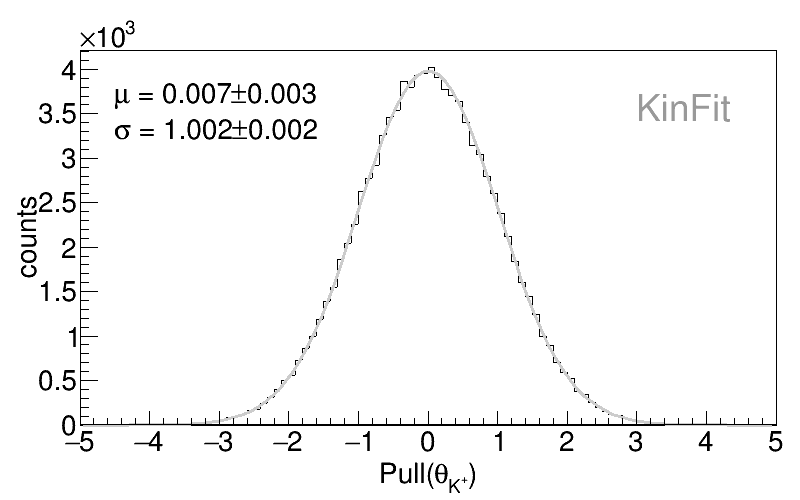}~
    \includegraphics[width= 0.34\textwidth]{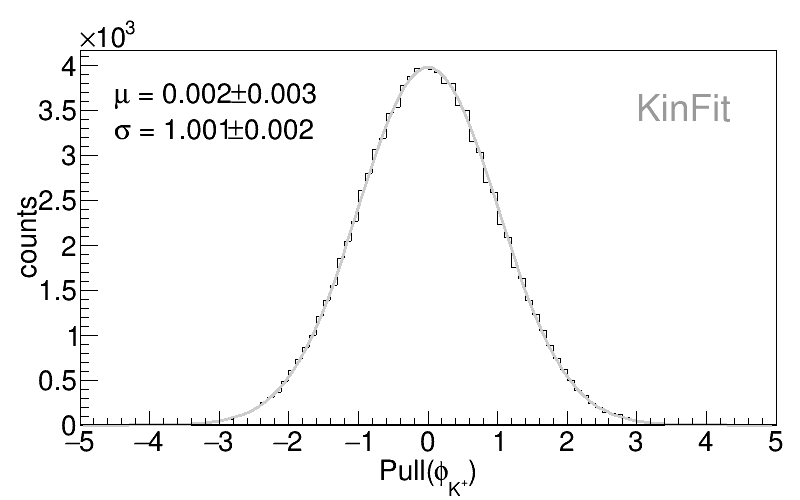}
    \includegraphics[width= 0.34\textwidth]{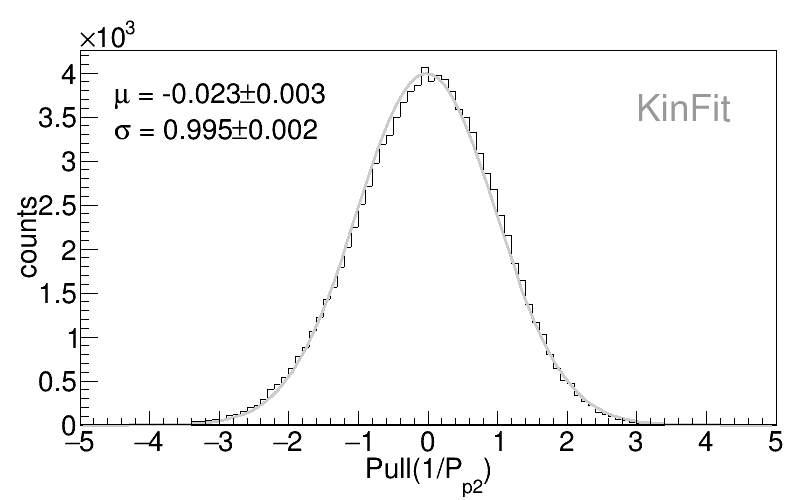}~
    \includegraphics[width= 0.34\textwidth]{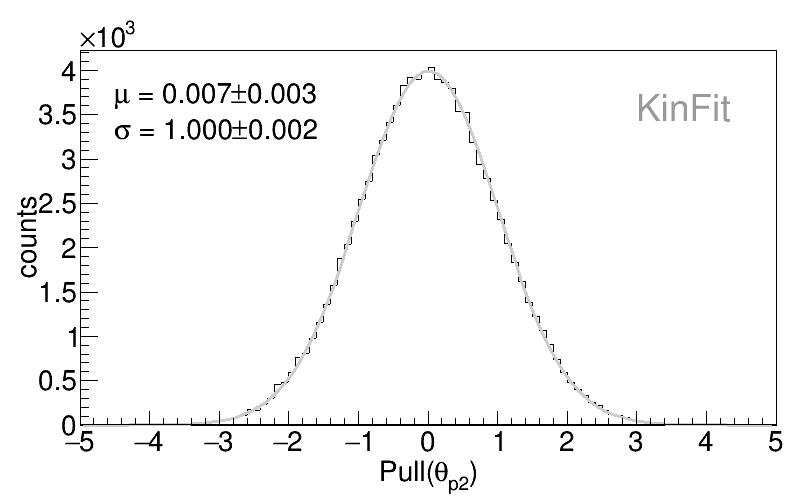}~
    \includegraphics[width= 0.34\textwidth]{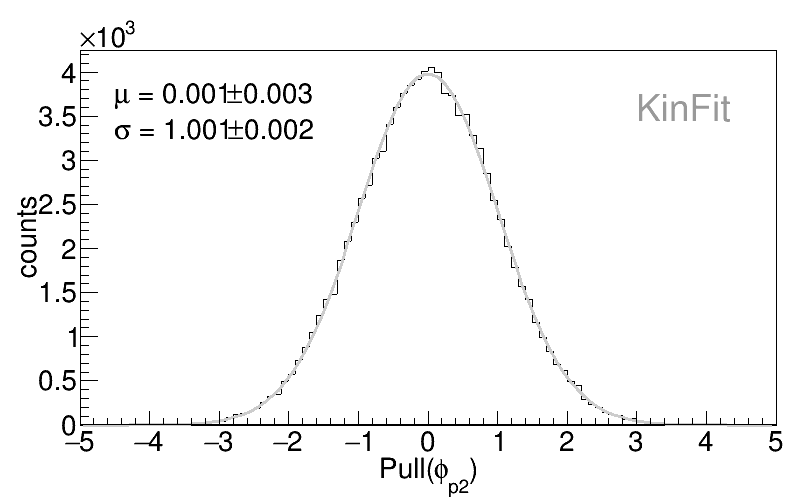}
    \includegraphics[width= 0.34\textwidth]{pics/pull_pi_p_4C}~
    \includegraphics[width= 0.34\textwidth]{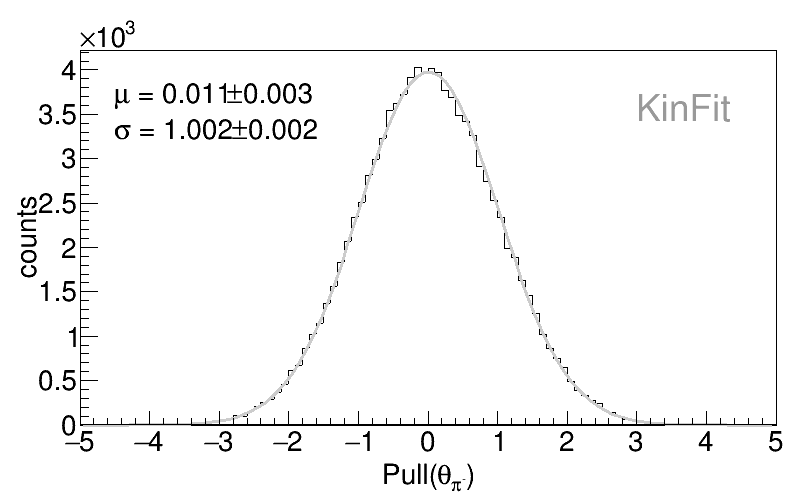}~
    \includegraphics[width= 0.34\textwidth]{pics/pull_pi_phi_4C}
    \caption{Pull distributions for the track parameters of all particles after the 4C fit with respective mean and standard deviation.}
    \label{fig:pulls_4C_extra}
\end{figure}
\pagebreak

\subsection{Pull Distributions of Missing Particle Fit}
\begin{figure}[!ht]
    \centering
    \includegraphics[width= 0.34\textwidth]{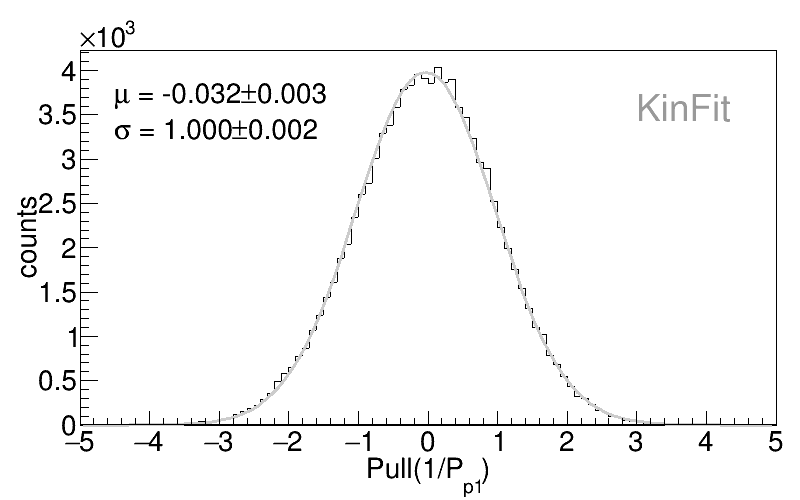}~
    \includegraphics[width= 0.34\textwidth]{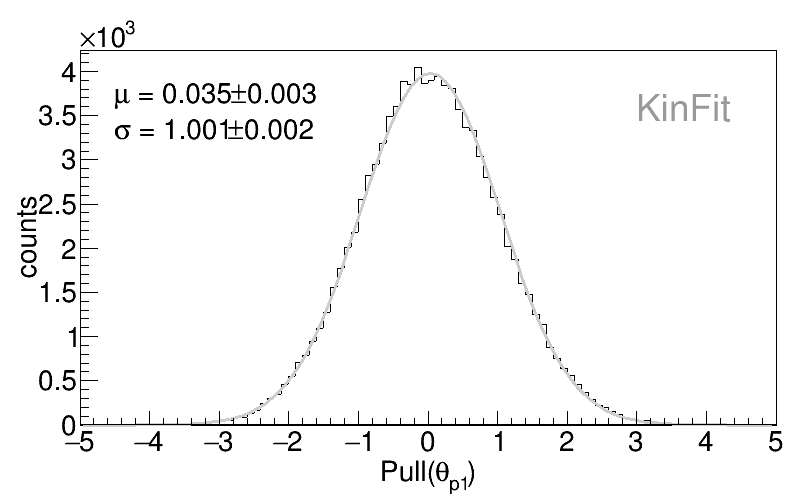}~
    \includegraphics[width= 0.34\textwidth]{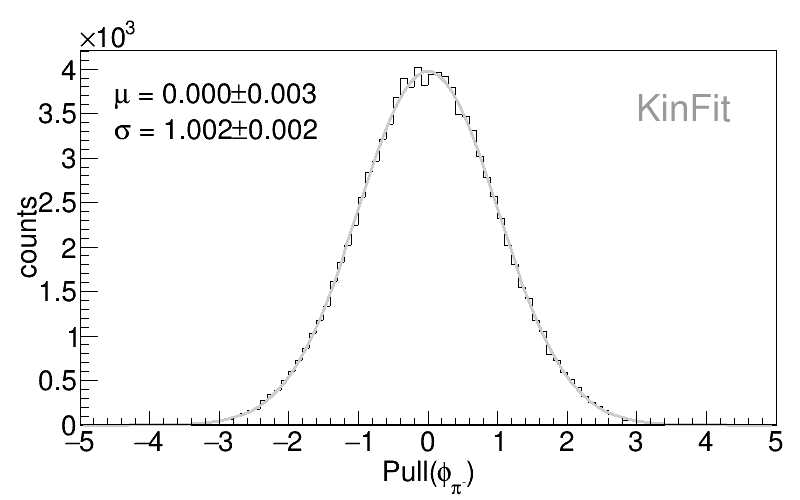}
    \includegraphics[width= 0.34\textwidth]{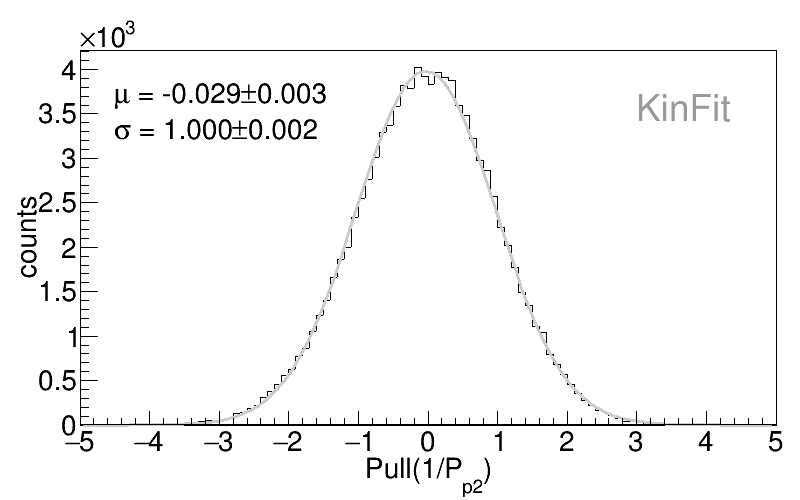}~
    \includegraphics[width= 0.34\textwidth]{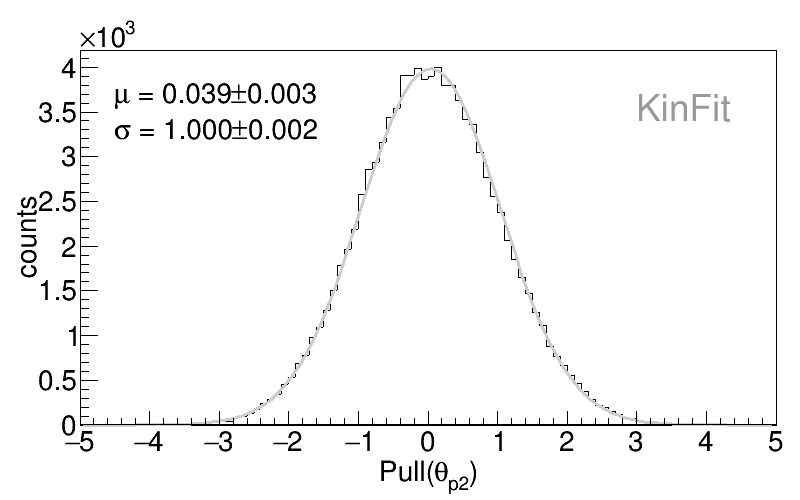}~
    \includegraphics[width= 0.34\textwidth]{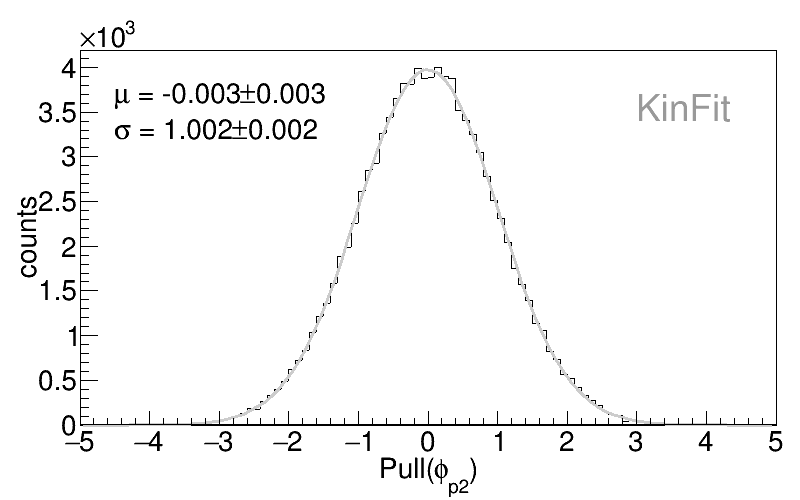}
    \includegraphics[width= 0.34\textwidth]{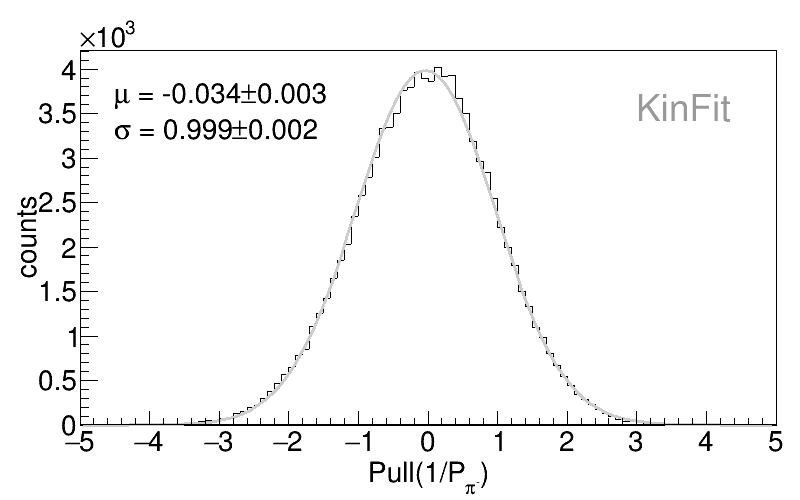}~
    \includegraphics[width= 0.34\textwidth]{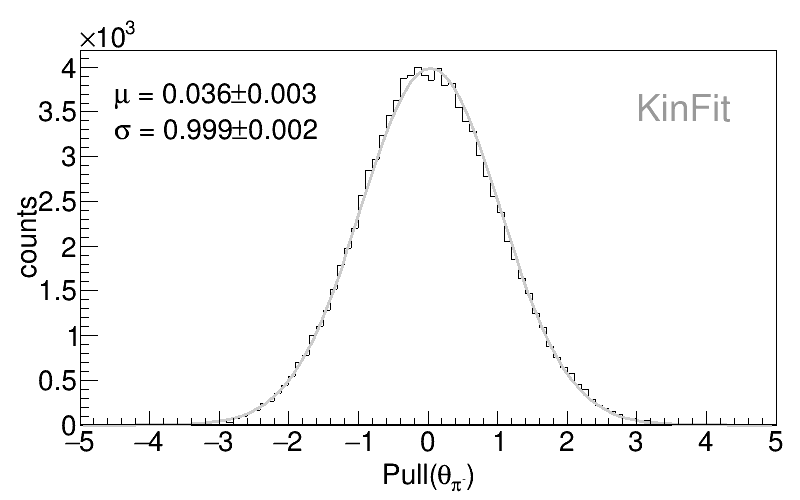}~
    \includegraphics[width= 0.34\textwidth]{pics/pull_pi_phi_miss}
    \caption{Pull distributions for the track parameters of all particles after the missing $K^+$ fit with respective mean and standard deviation.}
    \label{fig:pulls_miss_extra}
\end{figure}
\pagebreak

\subsection{Pull Distributions of 3C Fit in  \texorpdfstring{$\Lambda$}{lg} Decay Vertex}
\begin{figure}[!ht]
    \centering
    \includegraphics[width= 0.34\textwidth]{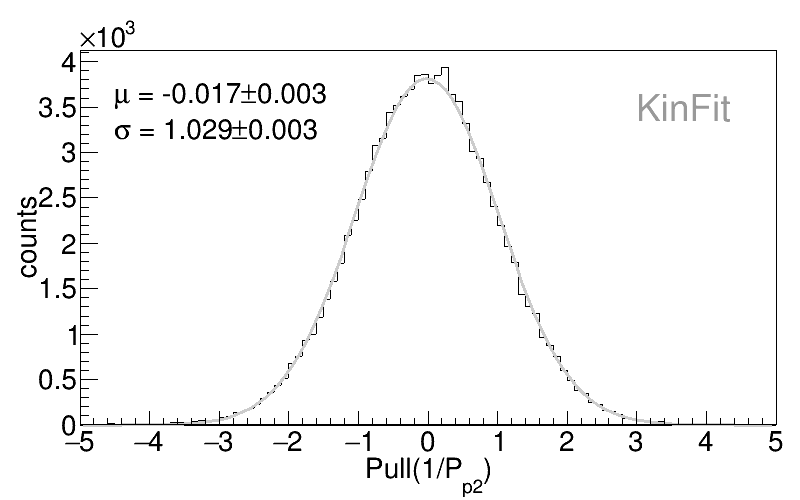 }~
    \includegraphics[width= 0.34\textwidth]{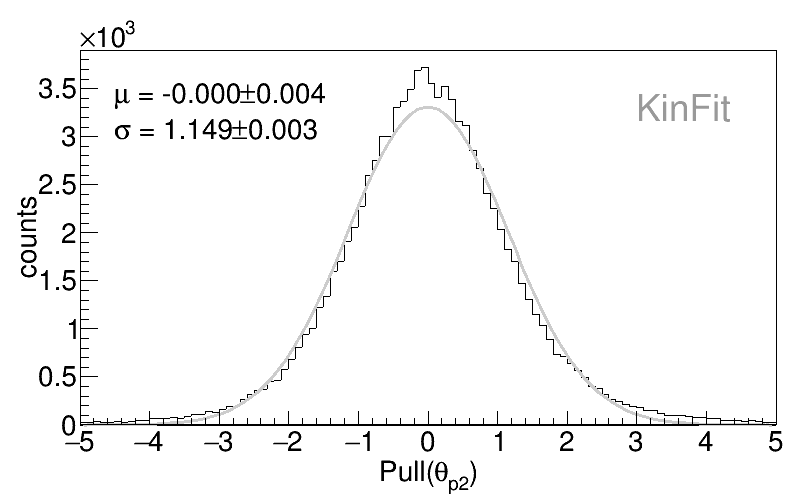 }~
    \includegraphics[width= 0.34\textwidth]{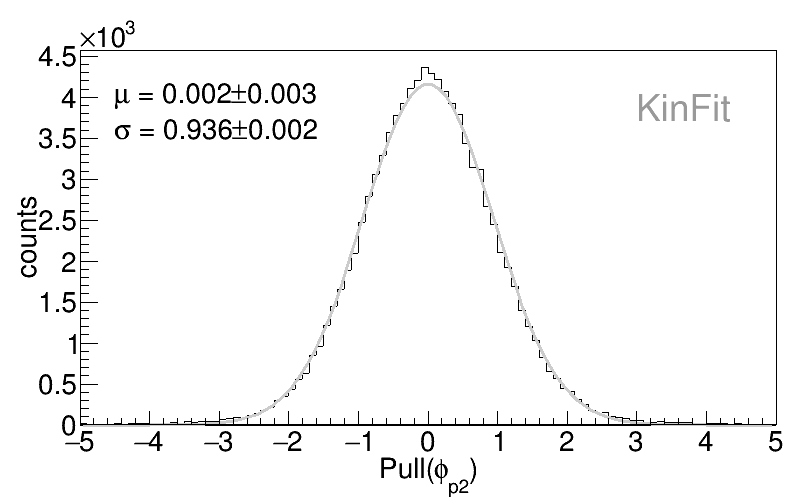 }
    \includegraphics[width= 0.34\textwidth]{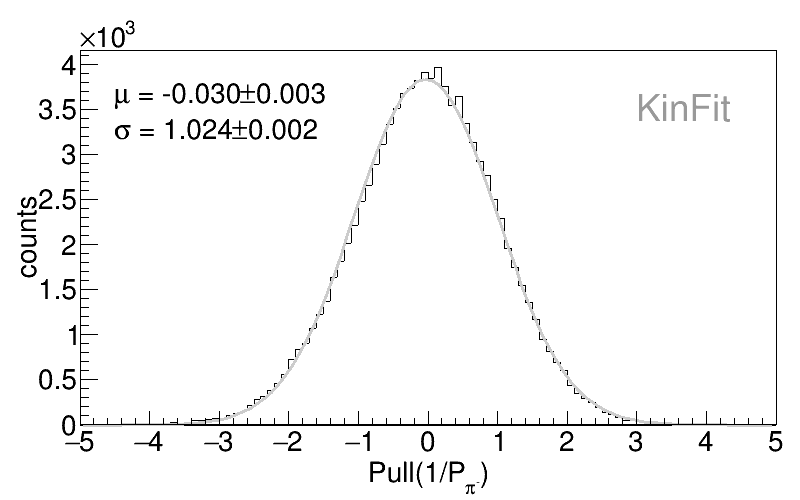 }~
    \includegraphics[width= 0.34\textwidth]{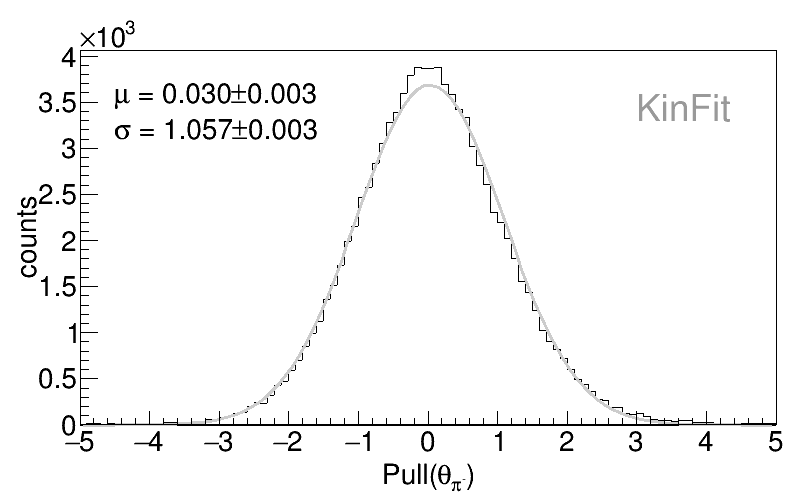 }~
    \includegraphics[width= 0.34\textwidth]{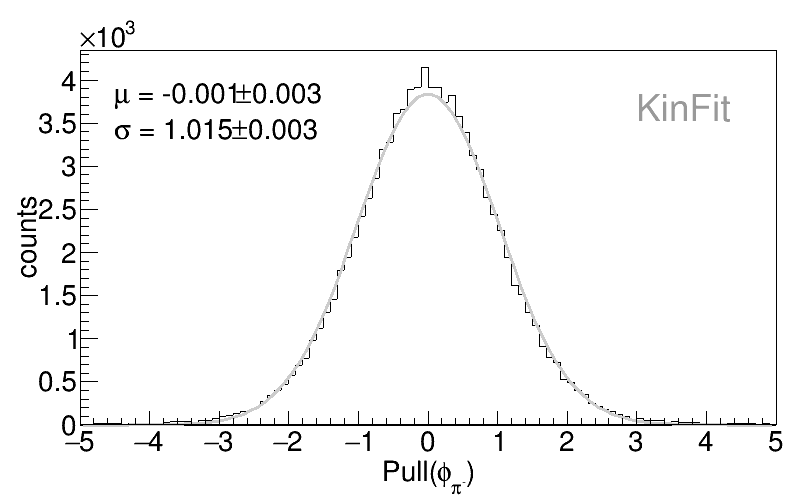 }
    \caption{Pull distributions for the track parameters of all particles after the 3C fit with respective mean and standard deviation.}
    \label{fig:pulls_3C_extra}
\end{figure}

\subsection{Pull Distributions of the Vertex Fit in the Interaction Point}
\begin{figure}[!ht]
    \centering
    \includegraphics[width= 0.34\textwidth]{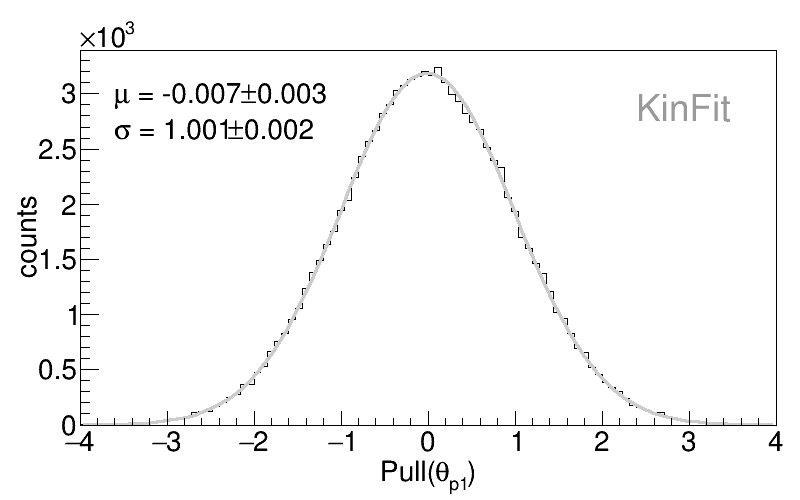 }~
    \includegraphics[width= 0.34\textwidth]{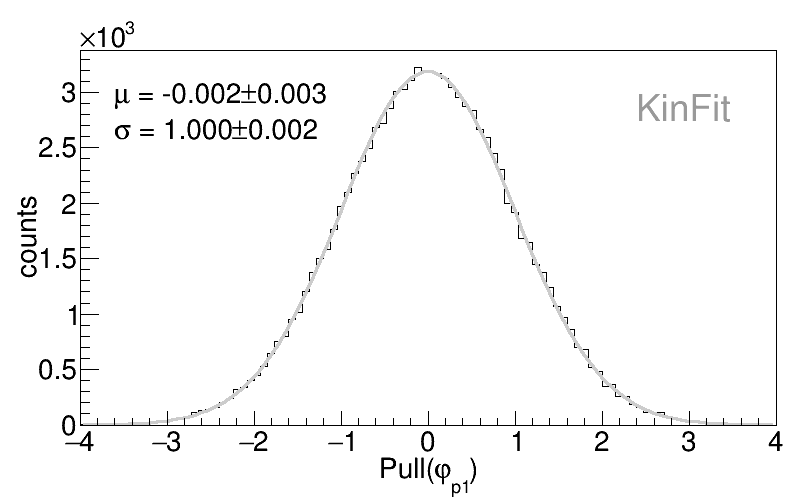 }~
    \includegraphics[width= 0.34\textwidth]{pics/vtx/p1_Rpull.png }
    \includegraphics[width= 0.34\textwidth]{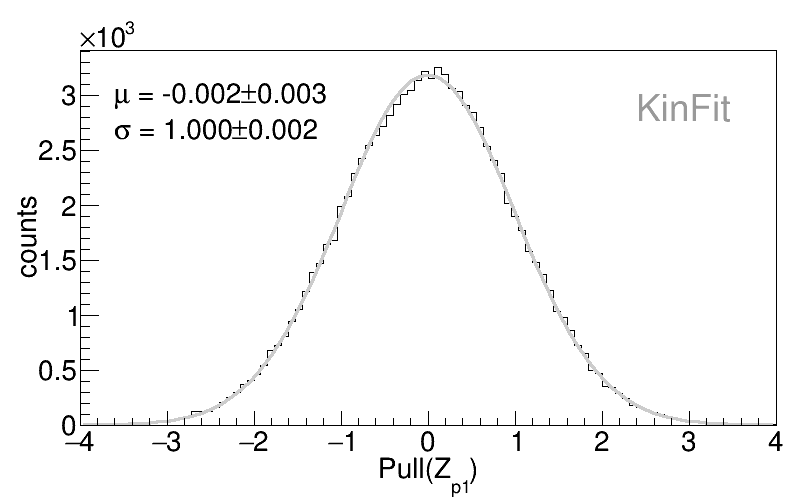}
    \caption{Pull distributions for the track parameters of the primary proton tracks after the vertex fit with respective mean and standard deviation.}
    \label{fig:pulls_vtx_extra}
\end{figure}


\end{document}